\documentclass[10pt, a4paper, conference]{IEEEtran}

\ifx\pdfoutput\undefined
\usepackage{graphicx}
\graphicspath{{eps_figs/}}
\else
\usepackage[pdftex]{graphicx}
\graphicspath{{pdf_figs/}}
\fi

\usepackage{latexsym,amssymb,amsmath,ifthen,afterpage,cite}

\usepackage{lipsum}
\usepackage{subcaption}
\usepackage{url}
\usepackage{tikz} 
\usepackage{pgfplots}
\usepackage{pgfplotstable}
\usetikzlibrary{plotmarks}
\pgfplotsset{compat = newest}
\pgfkeys{/pgf/number format/set thousands separator = }

\newcommand{\Fig}[1]{Fig.~\ref{#1}}

\newcommand{\eqdef}{\stackrel{\scriptscriptstyle\bigtriangleup}{=} }

\newcommand{\R}{\mathbb{R}}

\newcommand{\B}{{\mathcal{B}}}

\newcommand{\calX}{\mathcal{X}}

\newcommand{\calU}{\mathcal{U}}

\newcommand{\X}{\bf{X}}
\newcommand{\x}{{\bf x}}
\newcommand{\y}{{\bf y}}
\newcommand{\z}{{\bf z}}

\newcommand{\E}{\operatorname{E}}

\newcounter{examplecntr}
{\begin{trivlist}\small\item[]\refstepcounter{examplecntr}%
 {\bfseries Example~\theexamplecntr%
  \ifthenelse{\equal{#1}{}}{}{ (#1)}.
}}%
{\end{trivlist}}

\newcounter{theoremcntr}
{\begin{trivlist}\item[]\refstepcounter{theoremcntr}%
{\bfseries Theorem~\thetheoremcntr%
  \ifthenelse{\equal{#1}{}}{}{ (#1)}.
}}%
{\hfill$\Box$\end{trivlist}}

\newcommand{\pos}[2]{\makebox(0,0)[#1]{#2}}

  \renewenvironment{thebibliography}[1]{%
    \begin{oldthebibliography}{#1}%
      \setlength{\itemsep}{0.64ex}%
  }%
  {%
    \end{oldthebibliography}%
  }
  
\begin{document}
\DeclareGraphicsExtensions{.pdf}

\title{An Importance Sampling Algorithm\\ for the Ising Model with Strong Couplings\\} 

\author{Mehdi Molkaraie \\
ETH Zurich\\ 
8092 Z\"urich, Switzerland 
\\
\tt{mehdi.molkaraie@alumni.ethz.ch}\\
} 
 
\maketitle 
 
\begin{abstract} 
We consider the problem of estimating the partition function
of the ferromagnetic Ising model in a consistent external magnetic field. 
The estimation is done via
importance sampling in the dual of the Forney factor graph
representing the model. 
Emphasis is on models at low temperature (corresponding to models 
with strong couplings) and on models with a mixture of strong and weak
coupling parameters.
%
\end{abstract} 
 
\section{Introduction} 

The problem of estimating the partition function of the finite-size
two-dimensional (2D) 
ferromagnetic Ising model in a consistent external 
field is considered.
Applying factor graph duality to address the problem has been investigated 
in~\cite{MoLo:ISIT2013, MeMo:2014a, AY:2014, MeMo:2014b}. It was demonstrated in~\cite{MoLo:ISIT2013} that 
Monte Carlo methods based on the dual factor graph work very well for the Ising model at low temperature. In 
contrast, Monte Carlo methods 
in the primal/original 
graph suffer from critical slowing down and erratic convergence to estimate the
partition function in the low-temperature regime~\cite{BH:10}.
Monte Carlo methods (based on uniform sampling and Gibbs sampling) 
in the dual factor graph were also
proposed in~\cite{MoLo:ISIT2013}
to estimate the partition function of the 2D Ising model 
without an external field. 

In this paper, we continue this research to extend the 
results of~\cite{MoLo:ISIT2013, MeMo:2014a} to models with a mixture of strong and weak
coupling parameters and in the presence of an external magnetic field.
After defining an auxiliary probability mass function in 
the dual Forney factor graph of the
model, we propose an importance sampling algorithm that can
efficiently estimate the partition function.
A similar 
importance sampling algorithm, designed specifically for models in a strong external
field, was recently proposed in~\cite{MeMo:2014a}.

The paper is organized as follows. We review the Forney factor graph representation of the 2D Ising model 
in an external field in Section~\ref{sec:Ising}. Section~\ref{sec:NFGD} discusses
dual Forney factor graphs and 
the normal factor graph duality theorem. The importance sampling algorithm is described in Section~\ref{sec:IS}. 
In Section~\ref{sec:Num}, we report numerical experiments.

\section{The Ising Model in an External Magnetic Field}
\label{sec:Ising}

Let $X_1, X_2, \ldots, X_N$ be a set of discrete binary random variables
arranged on the sites of a 2D 
lattice. We suppose that 
interactions are restricted to 
adjacent (nearest-neighbor) variables (see Fig.~\ref{fig:2DGrid}). The real coupling parameter $J_{k, \ell}$ controls the strength of
the interaction between adjacent variables $(X_k, X_{\ell})$.
The real parameter $H_m$ corresponds to the
presence of an external field and controls the strength of
the interaction between $X_m$ and the field.
Each random
variable takes on values in $\calX = \{0, 1\}$.
Let $x_i$ represent a
possible realization of $X_i$, $\x$ stand for 
a configuration $(x_1, x_2, \ldots, x_N)$, and $\X$ stand for 
$(X_1, X_2, \ldots, X_N)$. 

The energy (the Hamiltonian)
of a configuration $\x$ is given by~\cite{Baxter07}
\begin{multline}
\label{eqn:HamiltonianI}
\mathcal{H}(\x) \, = -\!\!\sum_{\text{$(k,\ell)\in \B$}}\!\!\!J_{k, \ell}\cdot\big([x_k = x_{\ell}] - [x_k \ne x_{\ell}]\big)\\ 
- \sum_{m = 1}^N H_m\cdot\big([x_m = 1] - [x_m = 0]\big)
\end{multline}
where $\B$ contains 
all the unordered pairs (bonds) $(k,\ell)$ with non-zero 
interactions, and $[\cdot]$ denotes the Iverson 
bracket~\cite{Knuth:92}, which evaluates to $1$ if the condition in 
the bracket is satisfied and to $0$ otherwise.

In this paper, the focus is on ferromagnetic Ising models
characterized by  $J_{k, \ell} > 0$ for each $(k, \ell) \in \B$. The
external field is assumed to be consistent, i.e., it is either assigned to
all positive or to all negative values.


The probability that the model is in configuration $\x$ is
given by the Boltzmann distribution~\cite{Baxter07}
\begin{equation} 
\label{eqn:Prob}
p_{\text{B}}(\x) = \frac{e^{-\beta \mathcal{H}(\x)}}{Z} 
\end{equation}
where the normalization constant $Z$ is the \emph{partition function} 
$Z = \sum_{\x \in \calX^N} e^{-\beta \mathcal{H}(\x)}$ and
$\beta$ is the inverse temperature. 
In the rest of this paper, we assume $\beta = 1$. 
With this assumption, large values of $J$ correspond to models at low temperature. Boundary conditions are assumed to be periodic.



For each adjacent pair $(x_k, x_\ell)$, let $\kappa: \calX^2 \rightarrow \R_{>0}$
\begin{equation}
\label{eqn:IsingA}
\kappa_{k, \ell}(x_k, x_{\ell}) 
= e^{J_{k, \ell}\cdot\big([x_k = x_{\ell}] - [x_k \ne x_{\ell}]\big)}
\end{equation}
and for each $x_m$, let $\tau: \calX \rightarrow \R_{>0}$
\begin{equation}
\label{eqn:IsingH}
\tau_{m}(x_m) = e^{H_m\cdot\big([x_m = 1] - [x_m = 0]\big)}
\end{equation}

We then define $f: \calX^N \rightarrow \R_{>0}$ as 
\begin{equation} 
\label{eqn:factorF}
f(\x) \, \eqdef  \!\!\prod_{\text{$(k,\ell)\in \B$}}\!\!\!\kappa_{k, \ell}(x_k, x_{\ell})
 \prod_{m = 1}^N \tau_{m}(x_m)
\end{equation}

The corresponding Forney factor graph (normal graph) for the factorization
in~(\ref{eqn:factorF}) is shown in~\Fig{fig:2DGrid},
where the boxes labeled ``$=$'' are equality 
constraints~\cite{Forney:01, Lg:ifg2004}. In Forney factor graphs variables are 
represented by edges. 

From (\ref{eqn:factorF}), $Z$ in (\ref{eqn:Prob})
can also be expressed as 
\begin{equation}
\label{eqn:PartFunction}
Z = \sum_{\x \in \calX^N} f(\x) 
\end{equation}

At high temperature (i.e., for small $J$), 
the Boltzmann distribution~(\ref{eqn:Prob})
approaches the uniform distribution. In this case, Monte Carlo methods for estimating $Z$
usually perform well  
in the primal factor graph. Estimating $Z$ in the low-temperature regime is more challenging~\cite{HH:64,Neal:proinf1993r,BH:10}.

In this paper, we consider models at low temperature (i.e., with large $J$) and models with a mixture of
strong and weak coupling parameters in an external magnetic field.
To compute an estimate of $Z$ in this case, we propose an importance sampling algorithm in 
the dual of the Forney factor graph of the 2D Ising model.

\section{The Dual Forney Factor Graph}
\label{sec:NFGD}

We can obtain the dual of the Forney factor graph in~\Fig{fig:2DGrid}, by 
replacing each binary variable $x$ with its dual
binary variable $\tilde x$, each 
factor $\kappa_{k, \ell}$ with its 2D Discrete Fourier 
transform (DFT)\footnote{Here, $\gamma(\tilde x_1, \tilde x_2)$, the 2D DFT of $\kappa(x_1, x_2)$, 
is defined as
\begin{equation*} 
\gamma(\tilde x_1, \tilde x_2) \eqdef 
\sum_{x_1\in \calX}\sum_{x_2 \in \calX} \kappa(x_1,x_2)
e^{-i2\pi(x_1\tilde x_1 + x_2\tilde x_2)/|\calX|}
\end{equation*}
where $i $ is the unit imaginary number~\cite{Brace:1999}},
each factor $\tau_m$ with its one-dimensional (1D) 
DFT, and each equality 
constraint with an XOR factor, see~\cite{Forney:01, Forney:11, AY:2011, FV:2011}. 
\Fig{fig:2DGridDM} shows the dual Forney factor 
graph of the 2D Ising model, 
where boxes containing 
$``+"$ symbols represent XOR factors as 
\begin{equation} 
\label{eqn:XOR}
g(\tilde x_1, \tilde x_2, \ldots, \tilde x_k) =
[\tilde x_1 \oplus \tilde x_2 \oplus \ldots \oplus \tilde x_k=0]
\end{equation}
the small boxes attached to each
XOR factor are given by
\begin{equation} 
\label{eqn:IsingKernelDual2}
\lambda_{m}(\tilde x_m) = \left\{ \begin{array}{lr}
     \cosh H_{m}, & \text{if $\tilde x_m = 0$} \\
     -\sinh H_{m}, & \text{if $\tilde x_m = 1$} 
\end{array} \right.
\end{equation}
%
and the unlabeled 
normal-size boxes attached to each equality constraint represent 
factors as 
\begin{equation} 
\label{eqn:IsingKernelDual}
\gamma_{k}(\tilde x_k) = \left\{ \begin{array}{ll}
     2\cosh J_{k}, & \text{if $\tilde x_k = 0$} \\
     2\sinh J_{k}, & \text{if $\tilde x_k = 1$}
  \end{array} \right.
\end{equation}

Here, $J_k$ is 
the coupling parameter associated with each bond. 
See~\cite{MoLo:ISIT2013, MeMo:2014a, AY:2014}, for more details on constructing 
the dual Forney factor graph of the 2D Ising model. 

In the dual domain, 
we denote the partition function
by $Z_{\mathrm{d}}$. 
For the models that we study here,
the normal factor graph duality theorem states 
that 
\begin{equation}
\label{eqn:NDual}
Z_{\mathrm{d}} = |\calX^{N}|Z
\end{equation}
see~\cite[Theorem 2]{AY:2011}.

In order to design Monte Carlo methods in the dual Forney 
graph, we require factors~(\ref{eqn:IsingKernelDual2}) and~(\ref{eqn:IsingKernelDual}) to be non-negative.
In a 2D Ising model, $Z$ is invariant under the
change of sign of the external field~\cite{Baxter07}. Therefore, without
loss of generality,  
we will assume $H_m <  0$ for $1 \le m \le N$. 
Under the ferromagnetic assumption $J_{k, \ell} > 0$ for $(k, \ell) \in \B$.
With these assumptions, (\ref{eqn:IsingKernelDual2}) and~(\ref{eqn:IsingKernelDual}) will be non-negative. 

%

\begin{figure}[t]
\setlength{\unitlength}{0.88mm}
\centering
\begin{picture}(85,69.75)(0,0)
\small
\put(0,60){\framebox(4,4){$=$}}
 \put(4,60){\line(4,-3){4}}
 \put(8,54){\framebox(3,3){}}
\put(4,62){\line(1,0){8}}        \put(8,63){\pos{bc}{$X_1$}}
\put(12,60){\framebox(4,4){}}
\put(16,62){\line(1,0){8}}
\put(24,60){\framebox(4,4){$=$}}
 \put(28,60){\line(4,-3){4}}
 \put(32,54){\framebox(3,3){}}
\put(28,62){\line(1,0){8}}       \put(32,63){\pos{bc}{$X_2$}}
\put(36,60){\framebox(4,4){}}
\put(40,62){\line(1,0){8}}
\put(48,60){\framebox(4,4){$=$}}
 \put(52,60){\line(4,-3){4}}
 \put(56,54){\framebox(3,3){}}
\put(52,62){\line(1,0){8}}       
\put(60,60){\framebox(4,4){}}
\put(64,62){\line(1,0){8}}
\put(72,60){\framebox(4,4){$=$}}
 \put(76,60){\line(4,-3){4}}
 \put(80,54){\framebox(3,3){}}
\put(2,54){\line(0,1){6}}
\put(0,50){\framebox(4,4){}}
\put(2,50){\line(0,-1){6}}
\put(26,54){\line(0,1){6}}
\put(24,50){\framebox(4,4){}}
\put(26,50){\line(0,-1){6}}
\put(50,54){\line(0,1){6}}
\put(48,50){\framebox(4,4){}}
\put(50,50){\line(0,-1){6}}
\put(74,54){\line(0,1){6}}
\put(72,50){\framebox(4,4){}}
\put(74,50){\line(0,-1){6}}
\put(0,40){\framebox(4,4){$=$}}
 \put(4,40){\line(4,-3){4}}
 \put(8,34){\framebox(3,3){}}
\put(4,42){\line(1,0){8}}
\put(12,40){\framebox(4,4){}}
\put(16,42){\line(1,0){8}}
\put(24,40){\framebox(4,4){$=$}}
 \put(28,40){\line(4,-3){4}}
 \put(32,34){\framebox(3,3){}}
\put(28,42){\line(1,0){8}}
\put(36,40){\framebox(4,4){}}
\put(40,42){\line(1,0){8}}
\put(48,40){\framebox(4,4){$=$}}
 \put(52,40){\line(4,-3){4}}
 \put(56,34){\framebox(3,3){}}
\put(52,42){\line(1,0){8}}
\put(60,40){\framebox(4,4){}}
\put(64,42){\line(1,0){8}}
\put(72,40){\framebox(4,4){$=$}}
 \put(76,40){\line(4,-3){4}}
 \put(80,34){\framebox(3,3){}}
\put(2,34){\line(0,1){6}}
\put(0,30){\framebox(4,4){}}
\put(2,30){\line(0,-1){6}}
\put(26,34){\line(0,1){6}}
\put(24,30){\framebox(4,4){}}
\put(26,30){\line(0,-1){6}}
\put(50,34){\line(0,1){6}}
\put(48,30){\framebox(4,4){}}
\put(50,30){\line(0,-1){6}}
\put(74,34){\line(0,1){6}}
\put(72,30){\framebox(4,4){}}
\put(74,30){\line(0,-1){6}}
\put(0,20){\framebox(4,4){$=$}}
 \put(4,20){\line(4,-3){4}}
 \put(8,14){\framebox(3,3){}}
\put(4,22){\line(1,0){8}}
\put(12,20){\framebox(4,4){}}
\put(16,22){\line(1,0){8}}
\put(24,20){\framebox(4,4){$=$}}
 \put(28,20){\line(4,-3){4}}
 \put(32,14){\framebox(3,3){}}
\put(28,22){\line(1,0){8}}
\put(36,20){\framebox(4,4){}}
\put(40,22){\line(1,0){8}}
\put(48,20){\framebox(4,4){$=$}}
 \put(52,20){\line(4,-3){4}}
 \put(56,14){\framebox(3,3){}}
\put(52,22){\line(1,0){8}}
\put(60,20){\framebox(4,4){}}
\put(64,22){\line(1,0){8}}
\put(72,20){\framebox(4,4){$=$}}
 \put(76,20){\line(4,-3){4}}
 \put(80,14){\framebox(3,3){}}
\put(2,14){\line(0,1){6}}
\put(0,10){\framebox(4,4){}}
\put(2,10){\line(0,-1){6}}
\put(26,14){\line(0,1){6}}
\put(24,10){\framebox(4,4){}}
\put(26,10){\line(0,-1){6}}
\put(50,14){\line(0,1){6}}
\put(48,10){\framebox(4,4){}}
\put(50,10){\line(0,-1){6}}
\put(74,14){\line(0,1){6}}
\put(72,10){\framebox(4,4){}}
\put(74,10){\line(0,-1){6}}
\put(0,0){\framebox(4,4){$=$}}
 \put(4,0){\line(4,-3){4}}
 \put(8,-6){\framebox(3,3){}}
\put(4,2){\line(1,0){8}}
\put(12,0){\framebox(4,4){}}
\put(16,2){\line(1,0){8}}
\put(24,0){\framebox(4,4){$=$}}
 \put(28,0){\line(4,-3){4}}
\put(32,-6){\framebox(3,3){}}
\put(28,2){\line(1,0){8}}
\put(36,0){\framebox(4,4){}}
\put(40,2){\line(1,0){8}}
\put(48,0){\framebox(4,4){$=$}}
 \put(52,0){\line(4,-3){4}}
 \put(56,-6){\framebox(3,3){}}
\put(52,2){\line(1,0){8}}
\put(60,0){\framebox(4,4){}}
\put(64,2){\line(1,0){8}}
\put(72,0){\framebox(4,4){$=$}}
 \put(76,0){\line(4,-3){4}}
 \put(80,-6){\framebox(3,3){}}
 \put(14.2,64.85){\pos{bc}{$\kappa_{1,2}$}}
 \put(13.5,54){\pos{bc}{$\tau_{1}$}}
\end{picture}
%
\vspace{3.2ex}
\caption{\label{fig:2DGrid}%
Forney factor graph of the 2D Ising model in 
an external field, where unlabeled normal-size 
boxes represent~(\ref{eqn:IsingA}), 
small boxes represent~(\ref{eqn:IsingH}), and 
boxes containing $``="$ symbols are equality constraints.
}
\vspace{2ex}
%
%
\setlength{\unitlength}{0.88mm}
\centering
\begin{picture}(77,72.3)(1,0)
\small
\put(0,60){\framebox(4,4){$+$}}
 \put(4,60){\line(4,-3){4}}
 \put(8,54){\framebox(3,3){}}
\put(4,62){\line(1,0){8}}        
\put(12,60){\framebox(4,4){$=$}}
\put(16,62){\line(1,0){8}}
\put(24,60){\framebox(4,4){$+$}}
 \put(28,60){\line(4,-3){4}}
 \put(32,54){\framebox(3,3){}}
\put(28,62){\line(1,0){8}}       
\put(36,60){\framebox(4,4){$=$}}
\put(40,62){\line(1,0){8}}
\put(48,60){\framebox(4,4){$+$}}
 \put(52,60){\line(4,-3){4}}
 \put(56,54){\framebox(3,3){}}
\put(52,62){\line(1,0){8}}       
\put(60,60){\framebox(4,4){$=$}}
\put(64,62){\line(1,0){8}}
\put(72,60){\framebox(4,4){$+$}}
 \put(76,60){\line(4,-3){4}}
 \put(80,54){\framebox(3,3){}}
\put(2,54){\line(0,1){6}}
\put(0,50){\framebox(4,4){$=$}}
\put(2,50){\line(0,-1){6}}
\put(26,54){\line(0,1){6}}
\put(24,50){\framebox(4,4){$=$}}
\put(26,50){\line(0,-1){6}}
\put(50,54){\line(0,1){6}}
\put(48,50){\framebox(4,4){$=$}}
\put(50,50){\line(0,-1){6}}
\put(74,54){\line(0,1){6}}
\put(72,50){\framebox(4,4){$=$}}
\put(74,50){\line(0,-1){6}}
\put(0,40){\framebox(4,4){$+$}}
 \put(4,40){\line(4,-3){4}}
 \put(8,34){\framebox(3,3){}}
\put(4,42){\line(1,0){8}}
\put(12,40){\framebox(4,4){$=$}}
\put(16,42){\line(1,0){8}}
\put(24,40){\framebox(4,4){$+$}}
 \put(28,40){\line(4,-3){4}}
 \put(32,34){\framebox(3,3){}}
\put(28,42){\line(1,0){8}}
\put(36,40){\framebox(4,4){$=$}}
\put(40,42){\line(1,0){8}}
\put(48,40){\framebox(4,4){$+$}}
 \put(52,40){\line(4,-3){4}}
 \put(56,34){\framebox(3,3){}}
\put(52,42){\line(1,0){8}}
\put(60,40){\framebox(4,4){$=$}}
\put(64,42){\line(1,0){8}}
\put(72,40){\framebox(4,4){$+$}}
 \put(76,40){\line(4,-3){4}}
 \put(80,34){\framebox(3,3){}}
\put(2,34){\line(0,1){6}}
\put(0,30){\framebox(4,4){$=$}}
\put(2,30){\line(0,-1){6}}
\put(26,34){\line(0,1){6}}
\put(24,30){\framebox(4,4){$=$}}
\put(26,30){\line(0,-1){6}}
\put(50,34){\line(0,1){6}}
\put(48,30){\framebox(4,4){$=$}}
\put(50,30){\line(0,-1){6}}
\put(74,34){\line(0,1){6}}
\put(72,30){\framebox(4,4){$=$}}
\put(74,30){\line(0,-1){6}}
\put(0,20){\framebox(4,4){$+$}}
 \put(4,20){\line(4,-3){4}}
 \put(8,14){\framebox(3,3){}}
\put(4,22){\line(1,0){8}}
\put(12,20){\framebox(4,4){$=$}}
\put(16,22){\line(1,0){8}}
\put(24,20){\framebox(4,4){$+$}}
 \put(28,20){\line(4,-3){4}}
 \put(32,14){\framebox(3,3){}}
\put(28,22){\line(1,0){8}}
\put(36,20){\framebox(4,4){$=$}}
\put(40,22){\line(1,0){8}}
\put(48,20){\framebox(4,4){$+$}}
 \put(52,20){\line(4,-3){4}}
 \put(56,14){\framebox(3,3){}}
\put(52,22){\line(1,0){8}}
\put(60,20){\framebox(4,4){$=$}}
\put(64,22){\line(1,0){8}}
\put(72,20){\framebox(4,4){$+$}}
 \put(76,20){\line(4,-3){4}}
 \put(80,14){\framebox(3,3){}}
\put(2,14){\line(0,1){6}}
\put(0,10){\framebox(4,4){$=$}}
\put(2,10){\line(0,-1){6}}
\put(26,14){\line(0,1){6}}
\put(24,10){\framebox(4,4){$=$}}
\put(26,10){\line(0,-1){6}}
\put(50,14){\line(0,1){6}}
\put(48,10){\framebox(4,4){$=$}}
\put(50,10){\line(0,-1){6}}
\put(74,14){\line(0,1){6}}
\put(72,10){\framebox(4,4){$=$}}
\put(74,10){\line(0,-1){6}}
\put(0,0){\framebox(4,4){$+$}}
 \put(4,0){\line(4,-3){4}}
 \put(8,-6){\framebox(3,3){}}
\put(4,2){\line(1,0){8}}
\put(12,0){\framebox(4,4){$=$}}
\put(16,2){\line(1,0){8}}
\put(24,0){\framebox(4,4){$+$}}
 \put(28,0){\line(4,-3){4}}
\put(32,-6){\framebox(3,3){}}
\put(28,2){\line(1,0){8}}
\put(36,0){\framebox(4,4){$=$}}
\put(40,2){\line(1,0){8}}
\put(48,0){\framebox(4,4){$+$}}
 \put(52,0){\line(4,-3){4}}
 \put(56,-6){\framebox(3,3){}}
\put(52,2){\line(1,0){8}}
\put(60,0){\framebox(4,4){$=$}}
\put(64,2){\line(1,0){8}}
\put(72,0){\framebox(4,4){$+$}}
 \put(76,0){\line(4,-3){4}}
 \put(80,-6){\framebox(3,3){}}
 
\put(8,63){\pos{bc}{$\tilde X_1$}}
\put(32,63){\pos{bc}{$\tilde X_2$}}

\put(14,64){\line(0,1){2}}
\put(38,64){\line(0,1){2}}
\put(62,64){\line(0,1){2}}
\put(12,66){\framebox(4,4){$$}}
\put(36,66){\framebox(4,4){$$}}
\put(60,66){\framebox(4,4){$$}}
\put(14,44){\line(0,1){2}}
\put(38,44){\line(0,1){2}}
\put(62,44){\line(0,1){2}}
\put(12,46){\framebox(4,4){$$}}
\put(36,46){\framebox(4,4){$$}}
\put(60,46){\framebox(4,4){$$}}
\put(14,24){\line(0,1){2}}
\put(38,24){\line(0,1){2}}
\put(62,24){\line(0,1){2}}
\put(12,26){\framebox(4,4){$$}}
\put(36,26){\framebox(4,4){$$}}
\put(60,26){\framebox(4,4){$$}}
\put(14,4){\line(0,1){2}}
\put(38,4){\line(0,1){2}}
\put(62,4){\line(0,1){2}}
\put(12,6){\framebox(4,4){$$}}
\put(36,6){\framebox(4,4){$$}}
\put(60,6){\framebox(4,4){$$}}
\put(0,52){\line(-1,0){2}}
\put(24,52){\line(-1,0){2}}
\put(48,52){\line(-1,0){2}}
\put(72,52){\line(-1,0){2}}
\put(-6,50){\framebox(4,4){$$}}
\put(18,50){\framebox(4,4){$$}}
\put(42,50){\framebox(4,4){$$}}
\put(66,50){\framebox(4,4){$$}}
\put(0,32){\line(-1,0){2}}
\put(24,32){\line(-1,0){2}}
\put(48,32){\line(-1,0){2}}
\put(72,32){\line(-1,0){2}}
\put(-6,30){\framebox(4,4){$$}}
\put(18,30){\framebox(4,4){$$}}
\put(42,30){\framebox(4,4){$$}}
\put(66,30){\framebox(4,4){$$}}
\put(0,12){\line(-1,0){2}}
\put(24,12){\line(-1,0){2}}
\put(48,12){\line(-1,0){2}}
\put(72,12){\line(-1,0){2}}
\put(-6,10){\framebox(4,4){$$}}
\put(18,10){\framebox(4,4){$$}}
\put(42,10){\framebox(4,4){$$}}
\put(66,10){\framebox(4,4){$$}}
 \put(18.75,66.75){\pos{bc}{$\gamma_{1}$}}
 \put(13.5,54){\pos{bc}{$\lambda_{1}$}}
\end{picture}
\vspace{3.2ex}
\caption{\label{fig:2DGridDM}%
The dual Forney factor graph of the 2D Ising model in an external 
field, where boxes containing 
$``+"$ symbols represent~(\ref{eqn:XOR}), small 
boxes represent~(\ref{eqn:IsingKernelDual2}), and 
unlabeled normal-size boxes 
represent~(\ref{eqn:IsingKernelDual}). 
}
\end{figure}

\section{The Importance Sampling Algorithm}
\label{sec:IS}

The importance sampling algorithm is described 
on~\Fig{fig:2DGridDM}.
We partition $\tilde \X$ into $\tilde \X_A$ and $\tilde \X_B$, with the condition that $\tilde \X_B$ is a 
linear combination (involving the XOR factors) of $\tilde \X_A$. 
In this set-up, a valid configuration in the dual factor graph can be created by assigning
values to $\tilde \X_A$, followed by computing $\tilde \X_B$ as a linear combination of $\tilde \X_A$.

\begin{figure}[t]
\setlength{\unitlength}{0.88mm}
\centering
\begin{picture}(77,71.6)(0,0)
\small
\put(0,60){\framebox(4,4){$+$}}
 \put(8,54){\framebox(3,3){}}
\put(12,60){\framebox(4,4){$=$}}
\put(24,60){\framebox(4,4){$+$}}
 \put(32,54){\framebox(3,3){}}
\put(36,60){\framebox(4,4){$=$}}
\put(48,60){\framebox(4,4){$+$}}
 \put(56,54){\framebox(3,3){}}
\put(60,60){\framebox(4,4){$=$}}
\put(72,60){\framebox(4,4){$+$}}
 \put(80,54){\framebox(3,3){}}
 \put(0,50){\framebox(4,4){$=$}}
\put(24,50){\framebox(4,4){$=$}}
\put(48,50){\framebox(4,4){$=$}}
\put(72,50){\framebox(4,4){$=$}}
\put(0,40){\framebox(4,4){$+$}}
 \put(8,34){\framebox(3,3){}}
\put(12,40){\framebox(4,4){$=$}}
\put(24,40){\framebox(4,4){$+$}}
 \put(32,34){\framebox(3,3){}}
\put(36,40){\framebox(4,4){$=$}}
\put(48,40){\framebox(4,4){$+$}}
 \put(56,34){\framebox(3,3){}}
\put(60,40){\framebox(4,4){$=$}}
\put(72,40){\framebox(4,4){$+$}}
 \put(80,34){\framebox(3,3){}}
\put(0,30){\framebox(4,4){$=$}}
\put(24,30){\framebox(4,4){$=$}}
\put(48,30){\framebox(4,4){$=$}}
\put(72,30){\framebox(4,4){$=$}}
\put(0,20){\framebox(4,4){$+$}}
 \put(8,14){\framebox(3,3){}}
\put(12,20){\framebox(4,4){$=$}}
\put(24,20){\framebox(4,4){$+$}}
 \put(32,14){\framebox(3,3){}}
\put(36,20){\framebox(4,4){$=$}}
\put(48,20){\framebox(4,4){$+$}}
 \put(56,14){\framebox(3,3){}}
\put(60,20){\framebox(4,4){$=$}}
\put(72,20){\framebox(4,4){$+$}}
 \put(80,14){\framebox(3,3){}}
\put(0,10){\framebox(4,4){$=$}}
\put(24,10){\framebox(4,4){$=$}}
\put(48,10){\framebox(4,4){$=$}}
\put(72,10){\framebox(4,4){$=$}}
\put(0,0){\framebox(4,4){$+$}}
 \put(8,-6){\framebox(3,3){}}
\put(12,0){\framebox(4,4){$=$}}
\put(24,0){\framebox(4,4){$+$}}
\put(32,-6){\framebox(3,3){}}
\put(36,0){\framebox(4,4){$=$}}
\put(48,0){\framebox(4,4){$+$}}
 \put(56,-6){\framebox(3,3){}}
\put(60,0){\framebox(4,4){$=$}}
\put(72,0){\framebox(4,4){$+$}}
 \put(80,-6){\framebox(3,3){}}

\put(12,66){\framebox(4,4){$$}}
\put(36,66){\framebox(4,4){$$}}
\put(60,66){\framebox(4,4){$$}}
\put(12,46){\framebox(4,4){$$}}
\put(36,46){\framebox(4,4){$$}}
\put(60,46){\framebox(4,4){$$}}
\put(12,26){\framebox(4,4){$$}}
\put(36,26){\framebox(4,4){$$}}
\put(60,26){\framebox(4,4){$$}}
\put(12,6){\framebox(4,4){$$}}
\put(36,6){\framebox(4,4){$$}}
\put(60,6){\framebox(4,4){$$}}
\put(-6,50){\framebox(4,4){$$}}
\put(18,50){\framebox(4,4){$$}}
\put(42,50){\framebox(4,4){$$}}
\put(66,50){\framebox(4,4){$$}}
\put(-6,30){\framebox(4,4){$$}}
\put(18,30){\framebox(4,4){$$}}
\put(42,30){\framebox(4,4){$$}}
\put(66,30){\framebox(4,4){$$}}
\put(-6,10){\framebox(4,4){$$}}
\put(18,10){\framebox(4,4){$$}}
\put(42,10){\framebox(4,4){$$}}
\put(66,10){\framebox(4,4){$$}}

\put(4,60){\line(4,-3){4}}
\put(4,60){\line(4,-3){4}}
\put(4,60){\line(4,-3){4}}
\put(4,60){\line(4,-3){4}}
\put(4,60){\line(4,-3){4}}
\put(4,60){\line(4,-3){4}}
\put(4.1,59.9){\line(4,-3){4}}
\put(4.1,60.1){\line(4,-3){4}}
\put(3.9,59.9){\line(4,-3){4}}
\put(4,60){\line(4,-3){4}}
\put(4,60){\line(4,-3){4}}
\put(4,60){\line(4,-3){4}}
\put(4.1,59.9){\line(4,-3){4}}
\put(4.1,60.1){\line(4,-3){4}}
\put(3.9,59.9){\line(4,-3){4}}
\put(28,60){\line(4,-3){4}}
\put(28,60){\line(4,-3){4}}
\put(28,60){\line(4,-3){4}}
\put(28,60){\line(4,-3){4}}
\put(28,60){\line(4,-3){4}}
\put(28,60){\line(4,-3){4}}
\put(28.1,59.9){\line(4,-3){4}}
\put(28.1,60.1){\line(4,-3){4}}
\put(27.9,59.9){\line(4,-3){4}}
\put(28,60){\line(4,-3){4}}
\put(28,60){\line(4,-3){4}}
\put(28,60){\line(4,-3){4}}
\put(28.1,59.9){\line(4,-3){4}}
\put(28.1,60.1){\line(4,-3){4}}
\put(27.9,59.9){\line(4,-3){4}}
 \put(52,60){\line(4,-3){4}}
 \put(52,60){\line(4,-3){4}}
 \put(52,60){\line(4,-3){4}}
 \put(52,60){\line(4,-3){4}}
 \put(52,60){\line(4,-3){4}}
 \put(52,60){\line(4,-3){4}}
 \put(52.1,59.9){\line(4,-3){4}}
 \put(52.1,60.1){\line(4,-3){4}}
 \put(51.9,59.9){\line(4,-3){4}}
 \put(52,60){\line(4,-3){4}}
 \put(52,60){\line(4,-3){4}}
 \put(52,60){\line(4,-3){4}}
 \put(52.1,59.9){\line(4,-3){4}}
 \put(52.1,60.1){\line(4,-3){4}}
 \put(51.9,59.9){\line(4,-3){4}}
 \put(76,60){\line(4,-3){4}}
 \put(76,60){\line(4,-3){4}}
 \put(76,60){\line(4,-3){4}}
\put(76,60){\line(4,-3){4}}
 \put(76,60){\line(4,-3){4}}
 \put(76,60){\line(4,-3){4}}
 \put(76.1,59.9){\line(4,-3){4}}
 \put(76.1,60.1){\line(4,-3){4}}
 \put(75.9,59.9){\line(4,-3){4}}
 \put(76,60){\line(4,-3){4}}
 \put(76,60){\line(4,-3){4}}
 \put(76,60){\line(4,-3){4}}
 \put(76.1,59.9){\line(4,-3){4}}
 \put(76.1,60.1){\line(4,-3){4}}
 \put(75.9,59.9){\line(4,-3){4}}
 \put(4,40){\line(4,-3){4}}
 \put(4,40){\line(4,-3){4}}
 \put(4,40){\line(4,-3){4}}
 \put(4,40){\line(4,-3){4}}
 \put(4,40){\line(4,-3){4}}
 \put(4,40){\line(4,-3){4}}
 \put(4.1,39.9){\line(4,-3){4}}
 \put(4.1,40.1){\line(4,-3){4}}
 \put(3.9,39.9){\line(4,-3){4}}
 \put(4,40){\line(4,-3){4}}
 \put(4,40){\line(4,-3){4}}
 \put(4,40){\line(4,-3){4}}
 \put(4.1,39.9){\line(4,-3){4}}
 \put(4.1,40.1){\line(4,-3){4}}
 \put(3.9,39.9){\line(4,-3){4}}
 \put(28,40){\line(4,-3){4}}
 \put(28,40){\line(4,-3){4}}
 \put(28,40){\line(4,-3){4}}
 \put(28,40){\line(4,-3){4}}
 \put(28,40){\line(4,-3){4}}
 \put(28,40){\line(4,-3){4}}
 \put(28.1,39.9){\line(4,-3){4}}
 \put(28.1,40.1){\line(4,-3){4}}
 \put(27.9,39.9){\line(4,-3){4}}
\put(28,40){\line(4,-3){4}}
 \put(28,40){\line(4,-3){4}}
 \put(28,40){\line(4,-3){4}}
 \put(28.1,39.9){\line(4,-3){4}}
 \put(28.1,40.1){\line(4,-3){4}}
 \put(27.9,39.9){\line(4,-3){4}}
 \put(52,40){\line(4,-3){4}}
 \put(52,40){\line(4,-3){4}}
 \put(52,40){\line(4,-3){4}}
\put(52,40){\line(4,-3){4}}
 \put(52,40){\line(4,-3){4}}
 \put(52,40){\line(4,-3){4}}
 \put(52.1,39.9){\line(4,-3){4}}
 \put(52.1,40.1){\line(4,-3){4}}
 \put(51.9,39.9){\line(4,-3){4}}
 \put(52,40){\line(4,-3){4}}
 \put(52,40){\line(4,-3){4}}
 \put(52,40){\line(4,-3){4}}
 \put(52.1,39.9){\line(4,-3){4}}
 \put(52.1,40.1){\line(4,-3){4}}
 \put(51.9,39.9){\line(4,-3){4}}
 \put(76,40){\line(4,-3){4}}
 \put(76,40){\line(4,-3){4}}
 \put(76,40){\line(4,-3){4}}
\put(76,40){\line(4,-3){4}}
 \put(76,40){\line(4,-3){4}}
 \put(76,40){\line(4,-3){4}}
 \put(76.1,39.9){\line(4,-3){4}}
 \put(76.1,40.1){\line(4,-3){4}}
 \put(75.9,39.9){\line(4,-3){4}}
\put(76,40){\line(4,-3){4}}
 \put(76,40){\line(4,-3){4}}
 \put(76,40){\line(4,-3){4}}
 \put(76.1,39.9){\line(4,-3){4}}
 \put(76.1,40.1){\line(4,-3){4}}
 \put(75.9,39.9){\line(4,-3){4}}
 \put(4,20){\line(4,-3){4}}
 \put(4,20){\line(4,-3){4}}
 \put(4,20){\line(4,-3){4}}
 \put(4,20){\line(4,-3){4}}
 \put(4,20){\line(4,-3){4}}
 \put(4,20){\line(4,-3){4}}
 \put(4.1,19.9){\line(4,-3){4}}
 \put(4.1,20.1){\line(4,-3){4}}
 \put(3.9,19.9){\line(4,-3){4}}
 \put(4,20){\line(4,-3){4}}
 \put(4,20){\line(4,-3){4}}
 \put(4.1,19.9){\line(4,-3){4}}
 \put(4.1,20.1){\line(4,-3){4}}
 \put(3.9,19.9){\line(4,-3){4}}
 \put(28,20){\line(4,-3){4}}
 \put(28,20){\line(4,-3){4}}
 \put(28,20){\line(4,-3){4}}
 \put(28,20){\line(4,-3){4}}
 \put(28,20){\line(4,-3){4}}
 \put(28,20){\line(4,-3){4}}
 \put(28.1,19.9){\line(4,-3){4}}
 \put(28.1,20.1){\line(4,-3){4}}
 \put(27.9,19.9){\line(4,-3){4}}
 \put(28,20){\line(4,-3){4}}
 \put(28,20){\line(4,-3){4}}
 \put(28,20){\line(4,-3){4}}
 \put(28.1,19.9){\line(4,-3){4}}
 \put(28.1,20.1){\line(4,-3){4}}
 \put(27.9,19.9){\line(4,-3){4}}
 \put(52,20){\line(4,-3){4}}
 \put(52,20){\line(4,-3){4}}
 \put(52,20){\line(4,-3){4}}
 \put(52,20){\line(4,-3){4}}
 \put(52,20){\line(4,-3){4}}
 \put(52,20){\line(4,-3){4}}
 \put(52.1,19.9){\line(4,-3){4}}
 \put(52.1,20.1){\line(4,-3){4}}
 \put(51.9,19.9){\line(4,-3){4}}
 \put(52,20){\line(4,-3){4}}
 \put(52,20){\line(4,-3){4}}
 \put(52,20){\line(4,-3){4}}
 \put(52.1,19.9){\line(4,-3){4}}
 \put(52.1,20.1){\line(4,-3){4}}
 \put(51.9,19.9){\line(4,-3){4}}
 \put(76,20){\line(4,-3){4}}
 \put(76,20){\line(4,-3){4}}
 \put(76,20){\line(4,-3){4}}
\put(76,20){\line(4,-3){4}}
 \put(76,20){\line(4,-3){4}}
 \put(76,20){\line(4,-3){4}}
 \put(76.1,19.9){\line(4,-3){4}}
 \put(76.1,20.1){\line(4,-3){4}}
 \put(75.9,19.9){\line(4,-3){4}}
\put(76,20){\line(4,-3){4}}
 \put(76,20){\line(4,-3){4}}
 \put(76,20){\line(4,-3){4}}
 \put(76.1,19.9){\line(4,-3){4}}
 \put(76.1,20.1){\line(4,-3){4}}
 \put(75.9,19.9){\line(4,-3){4}}
 \put(4,0){\line(4,-3){4}}
 \put(4,0){\line(4,-3){4}}
 \put(4,0){\line(4,-3){4}}
 \put(4,0){\line(4,-3){4}}
 \put(4,0){\line(4,-3){4}}
 \put(4,0){\line(4,-3){4}}
 \put(4.1,-0.1){\line(4,-3){4}}
 \put(4.1,0.1){\line(4,-3){4}}
 \put(3.9,-0.1){\line(4,-3){4}}
\put(4,0){\line(4,-3){4}}
 \put(4,0){\line(4,-3){4}}
 \put(4,0){\line(4,-3){4}}
 \put(4.1,-0.1){\line(4,-3){4}}
 \put(4.1,0.1){\line(4,-3){4}}
 \put(3.9,-0.1){\line(4,-3){4}}
 \put(28,0){\line(4,-3){4}}
 \put(28,0){\line(4,-3){4}}
 \put(28,0){\line(4,-3){4}}
\put(28,0){\line(4,-3){4}}
 \put(28,0){\line(4,-3){4}}
 \put(28,0){\line(4,-3){4}}
 \put(28.1,-0.1){\line(4,-3){4}}
 \put(28.1,0.1){\line(4,-3){4}}
 \put(27.9,-0.1){\line(4,-3){4}}
 \put(28,0){\line(4,-3){4}}
 \put(28,0){\line(4,-3){4}}
 \put(28,0){\line(4,-3){4}}
 \put(28.1,-0.1){\line(4,-3){4}}
 \put(28.1,0.1){\line(4,-3){4}}
 \put(27.9,-0.1){\line(4,-3){4}}
 \put(52,0){\line(4,-3){4}}
 \put(52,0){\line(4,-3){4}}
 \put(52,0){\line(4,-3){4}}
 \put(52,0){\line(4,-3){4}}
 \put(52,0){\line(4,-3){4}}
 \put(52,0){\line(4,-3){4}}
 \put(52.1,-0.1){\line(4,-3){4}}
 \put(52.1,0.1){\line(4,-3){4}}
 \put(51.9,-0.1){\line(4,-3){4}}
 \put(52,0){\line(4,-3){4}}
 \put(52,0){\line(4,-3){4}}
 \put(52,0){\line(4,-3){4}}
 \put(52.1,-0.1){\line(4,-3){4}}
 \put(52.1,0.1){\line(4,-3){4}}
 \put(51.9,-0.1){\line(4,-3){4}}
 \put(76,0){\line(4,-3){4}}
 \put(76,0){\line(4,-3){4}}
 \put(76,0){\line(4,-3){4}}
 \put(76,0){\line(4,-3){4}}
 \put(76,0){\line(4,-3){4}}
 \put(76,0){\line(4,-3){4}}
 \put(76.1,-0.1){\line(4,-3){4}}
 \put(76.1,0.1){\line(4,-3){4}}
 \put(75.9,-0.1){\line(4,-3){4}}
\put(76,0){\line(4,-3){4}}
 \put(76,0){\line(4,-3){4}}
 \put(76,0){\line(4,-3){4}}
 \put(76.1,-0.1){\line(4,-3){4}}
 \put(76.1,0.1){\line(4,-3){4}}
 \put(75.9,-0.1){\line(4,-3){4}}
\put(4,60){\line(4,-3){4}}
\put(4,60){\line(4,-3){4}}
\put(4,60){\line(4,-3){4}}
\put(4,60){\line(4,-3){4}}
\put(4,60){\line(4,-3){4}}
\put(4,60){\line(4,-3){4}}
\put(4.1,59.9){\line(4,-3){4}}
\put(4.1,60.1){\line(4,-3){4}}
\put(3.9,59.9){\line(4,-3){4}}
\put(4,60){\line(4,-3){4}}
\put(4,60){\line(4,-3){4}}
\put(4,60){\line(4,-3){4}}
\put(4.1,59.9){\line(4,-3){4}}
\put(4.1,60.1){\line(4,-3){4}}
\put(3.9,59.9){\line(4,-3){4}}
\put(28,60){\line(4,-3){4}}
\put(28,60){\line(4,-3){4}}
\put(28,60){\line(4,-3){4}}
\put(28,60){\line(4,-3){4}}
\put(28,60){\line(4,-3){4}}
\put(28,60){\line(4,-3){4}}
\put(28.1,59.9){\line(4,-3){4}}
\put(28.1,60.1){\line(4,-3){4}}
\put(27.9,59.9){\line(4,-3){4}}
\put(28,60){\line(4,-3){4}}
\put(28,60){\line(4,-3){4}}
\put(28,60){\line(4,-3){4}}
\put(28.1,59.9){\line(4,-3){4}}
\put(28.1,60.1){\line(4,-3){4}}
\put(27.9,59.9){\line(4,-3){4}}
 \put(52,60){\line(4,-3){4}}
 \put(52,60){\line(4,-3){4}}
 \put(52,60){\line(4,-3){4}}
 \put(52,60){\line(4,-3){4}}
 \put(52,60){\line(4,-3){4}}
 \put(52,60){\line(4,-3){4}}
 \put(52.1,59.9){\line(4,-3){4}}
 \put(52.1,60.1){\line(4,-3){4}}
 \put(51.9,59.9){\line(4,-3){4}}
 \put(52,60){\line(4,-3){4}}
 \put(52,60){\line(4,-3){4}}
 \put(52,60){\line(4,-3){4}}
 \put(52.1,59.9){\line(4,-3){4}}
 \put(52.1,60.1){\line(4,-3){4}}
 \put(51.9,59.9){\line(4,-3){4}}
 \put(76,60){\line(4,-3){4}}
 \put(76,60){\line(4,-3){4}}
 \put(76,60){\line(4,-3){4}}
\put(76,60){\line(4,-3){4}}
 \put(76,60){\line(4,-3){4}}
 \put(76,60){\line(4,-3){4}}
 \put(76.1,59.9){\line(4,-3){4}}
 \put(76.1,60.1){\line(4,-3){4}}
 \put(75.9,59.9){\line(4,-3){4}}
 \put(76,60){\line(4,-3){4}}
 \put(76,60){\line(4,-3){4}}
 \put(76,60){\line(4,-3){4}}
 \put(76.1,59.9){\line(4,-3){4}}
 \put(76.1,60.1){\line(4,-3){4}}
 \put(75.9,59.9){\line(4,-3){4}}
 \put(4,40){\line(4,-3){4}}
 \put(4,40){\line(4,-3){4}}
 \put(4,40){\line(4,-3){4}}
 \put(4,40){\line(4,-3){4}}
 \put(4,40){\line(4,-3){4}}
 \put(4,40){\line(4,-3){4}}
 \put(4.1,39.9){\line(4,-3){4}}
 \put(4.1,40.1){\line(4,-3){4}}
 \put(3.9,39.9){\line(4,-3){4}}
 \put(4,40){\line(4,-3){4}}
 \put(4,40){\line(4,-3){4}}
 \put(4,40){\line(4,-3){4}}
 \put(4.1,39.9){\line(4,-3){4}}
 \put(4.1,40.1){\line(4,-3){4}}
 \put(3.9,39.9){\line(4,-3){4}}
 \put(28,40){\line(4,-3){4}}
 \put(28,40){\line(4,-3){4}}
 \put(28,40){\line(4,-3){4}}
 \put(28,40){\line(4,-3){4}}
 \put(28,40){\line(4,-3){4}}
 \put(28,40){\line(4,-3){4}}
 \put(28.1,39.9){\line(4,-3){4}}
 \put(28.1,40.1){\line(4,-3){4}}
 \put(27.9,39.9){\line(4,-3){4}}
\put(28,40){\line(4,-3){4}}
 \put(28,40){\line(4,-3){4}}
 \put(28,40){\line(4,-3){4}}
 \put(28.1,39.9){\line(4,-3){4}}
 \put(28.1,40.1){\line(4,-3){4}}
 \put(27.9,39.9){\line(4,-3){4}}
 \put(52,40){\line(4,-3){4}}
 \put(52,40){\line(4,-3){4}}
 \put(52,40){\line(4,-3){4}}
\put(52,40){\line(4,-3){4}}
 \put(52,40){\line(4,-3){4}}
 \put(52,40){\line(4,-3){4}}
 \put(52.1,39.9){\line(4,-3){4}}
 \put(52.1,40.1){\line(4,-3){4}}
 \put(51.9,39.9){\line(4,-3){4}}
 \put(52,40){\line(4,-3){4}}
 \put(52,40){\line(4,-3){4}}
 \put(52,40){\line(4,-3){4}}
 \put(52.1,39.9){\line(4,-3){4}}
 \put(52.1,40.1){\line(4,-3){4}}
 \put(51.9,39.9){\line(4,-3){4}}
 \put(76,40){\line(4,-3){4}}
 \put(76,40){\line(4,-3){4}}
 \put(76,40){\line(4,-3){4}}
\put(76,40){\line(4,-3){4}}
 \put(76,40){\line(4,-3){4}}
 \put(76,40){\line(4,-3){4}}
 \put(76.1,39.9){\line(4,-3){4}}
 \put(76.1,40.1){\line(4,-3){4}}
 \put(75.9,39.9){\line(4,-3){4}}
\put(76,40){\line(4,-3){4}}
 \put(76,40){\line(4,-3){4}}
 \put(76,40){\line(4,-3){4}}
 \put(76.1,39.9){\line(4,-3){4}}
 \put(76.1,40.1){\line(4,-3){4}}
 \put(75.9,39.9){\line(4,-3){4}}
 \put(4,20){\line(4,-3){4}}
 \put(4,20){\line(4,-3){4}}
 \put(4,20){\line(4,-3){4}}
 \put(4,20){\line(4,-3){4}}
 \put(4,20){\line(4,-3){4}}
 \put(4,20){\line(4,-3){4}}
 \put(4.1,19.9){\line(4,-3){4}}
 \put(4.1,20.1){\line(4,-3){4}}
 \put(3.9,19.9){\line(4,-3){4}}
 \put(4,20){\line(4,-3){4}}
 \put(4,20){\line(4,-3){4}}
 \put(4.1,19.9){\line(4,-3){4}}
 \put(4.1,20.1){\line(4,-3){4}}
 \put(3.9,19.9){\line(4,-3){4}}
 \put(28,20){\line(4,-3){4}}
 \put(28,20){\line(4,-3){4}}
 \put(28,20){\line(4,-3){4}}
 \put(28,20){\line(4,-3){4}}
 \put(28,20){\line(4,-3){4}}
 \put(28,20){\line(4,-3){4}}
 \put(28.1,19.9){\line(4,-3){4}}
 \put(28.1,20.1){\line(4,-3){4}}
 \put(27.9,19.9){\line(4,-3){4}}
 \put(28,20){\line(4,-3){4}}
 \put(28,20){\line(4,-3){4}}
 \put(28,20){\line(4,-3){4}}
 \put(28.1,19.9){\line(4,-3){4}}
 \put(28.1,20.1){\line(4,-3){4}}
 \put(27.9,19.9){\line(4,-3){4}}
 \put(52,20){\line(4,-3){4}}
 \put(52,20){\line(4,-3){4}}
 \put(52,20){\line(4,-3){4}}
 \put(52,20){\line(4,-3){4}}
 \put(52,20){\line(4,-3){4}}
 \put(52,20){\line(4,-3){4}}
 \put(52.1,19.9){\line(4,-3){4}}
 \put(52.1,20.1){\line(4,-3){4}}
 \put(51.9,19.9){\line(4,-3){4}}
 \put(52,20){\line(4,-3){4}}
 \put(52,20){\line(4,-3){4}}
 \put(52,20){\line(4,-3){4}}
 \put(52.1,19.9){\line(4,-3){4}}
 \put(52.1,20.1){\line(4,-3){4}}
 \put(51.9,19.9){\line(4,-3){4}}
 \put(76,20){\line(4,-3){4}}
 \put(76,20){\line(4,-3){4}}
 \put(76,20){\line(4,-3){4}}
\put(76,20){\line(4,-3){4}}
 \put(76,20){\line(4,-3){4}}
 \put(76,20){\line(4,-3){4}}
 \put(76.1,19.9){\line(4,-3){4}}
 \put(76.1,20.1){\line(4,-3){4}}
 \put(75.9,19.9){\line(4,-3){4}}
\put(76,20){\line(4,-3){4}}
 \put(76,20){\line(4,-3){4}}
 \put(76,20){\line(4,-3){4}}
 \put(76.1,19.9){\line(4,-3){4}}
 \put(76.1,20.1){\line(4,-3){4}}
 \put(75.9,19.9){\line(4,-3){4}}
 \put(4,0){\line(4,-3){4}}
 \put(4,0){\line(4,-3){4}}
 \put(4,0){\line(4,-3){4}}
 \put(4,0){\line(4,-3){4}}
 \put(4,0){\line(4,-3){4}}
 \put(4,0){\line(4,-3){4}}
 \put(4.1,-0.1){\line(4,-3){4}}
 \put(4.1,0.1){\line(4,-3){4}}
 \put(3.9,-0.1){\line(4,-3){4}}
\put(4,0){\line(4,-3){4}}
 \put(4,0){\line(4,-3){4}}
 \put(4,0){\line(4,-3){4}}
 \put(4.1,-0.1){\line(4,-3){4}}
 \put(4.1,0.1){\line(4,-3){4}}
 \put(3.9,-0.1){\line(4,-3){4}}
 \put(28,0){\line(4,-3){4}}
 \put(28,0){\line(4,-3){4}}
 \put(28,0){\line(4,-3){4}}
\put(28,0){\line(4,-3){4}}
 \put(28,0){\line(4,-3){4}}
 \put(28,0){\line(4,-3){4}}
 \put(28.1,-0.1){\line(4,-3){4}}
 \put(28.1,0.1){\line(4,-3){4}}
 \put(27.9,-0.1){\line(4,-3){4}}
 \put(28,0){\line(4,-3){4}}
 \put(28,0){\line(4,-3){4}}
 \put(28,0){\line(4,-3){4}}
 \put(28.1,-0.1){\line(4,-3){4}}
 \put(28.1,0.1){\line(4,-3){4}}
 \put(27.9,-0.1){\line(4,-3){4}}
 \put(52,0){\line(4,-3){4}}
 \put(52,0){\line(4,-3){4}}
 \put(52,0){\line(4,-3){4}}
 \put(52,0){\line(4,-3){4}}
 \put(52,0){\line(4,-3){4}}
 \put(52,0){\line(4,-3){4}}
 \put(52.1,-0.1){\line(4,-3){4}}
 \put(52.1,0.1){\line(4,-3){4}}
 \put(51.9,-0.1){\line(4,-3){4}}
 \put(52,0){\line(4,-3){4}}
 \put(52,0){\line(4,-3){4}}
 \put(52,0){\line(4,-3){4}}
 \put(52.1,-0.1){\line(4,-3){4}}
 \put(52.1,0.1){\line(4,-3){4}}
 \put(51.9,-0.1){\line(4,-3){4}}
 \put(76,0){\line(4,-3){4}}
 \put(76,0){\line(4,-3){4}}
 \put(76,0){\line(4,-3){4}}
 \put(76,0){\line(4,-3){4}}
 \put(76,0){\line(4,-3){4}}
 \put(76,0){\line(4,-3){4}}
 \put(76.1,-0.1){\line(4,-3){4}}
 \put(76.1,0.1){\line(4,-3){4}}
 \put(75.9,-0.1){\line(4,-3){4}}
\put(76,0){\line(4,-3){4}}
 \put(76,0){\line(4,-3){4}}
 \put(76,0){\line(4,-3){4}}
 \put(76.1,-0.1){\line(4,-3){4}}
 \put(76.1,0.1){\line(4,-3){4}}
 \put(75.9,-0.1){\line(4,-3){4}}
%
\put(2,54){\line(0,1){6}}
\put(2,50){\line(0,-1){6}}
\put(26,54){\line(0,1){6}}
\put(26,50){\line(0,-1){6}}
\put(50,54){\line(0,1){6}}
\put(50,50){\line(0,-1){6}}
\put(74,54){\line(0,1){6}}
\put(74,50){\line(0,-1){6}}
\put(2,34){\line(0,1){6}}
\put(2,30){\line(0,-1){6}}
\put(26,34){\line(0,1){6}}
\put(26,30){\line(0,-1){6}}
\put(50,34){\line(0,1){6}}
\put(50,30){\line(0,-1){6}}
\put(74,34){\line(0,1){6}}
\put(74,30){\line(0,-1){6}}
\put(2,14){\line(0,1){6}}
\put(2,10){\line(0,-1){6}}
\put(26,14){\line(0,1){6}}
\put(26,10){\line(0,-1){6}}
\put(50,14){\line(0,1){6}}
\put(50,10){\line(0,-1){6}}
\put(74,14){\line(0,1){6}}
\put(74,10){\line(0,-1){6}}
\put(4,2){\line(1,0){8}}
\put(16,2){\line(1,0){8}}
\put(28,2){\line(1,0){8}}
\put(40,2){\line(1,0){8}}
\put(52,2){\line(1,0){8}}
\put(64,2){\line(1,0){8}}
\put(0,52){\line(-1,0){2}}
\put(24,52){\line(-1,0){2}}
\put(48,52){\line(-1,0){2}}
\put(72,52){\line(-1,0){2}}

\put(0,32){\line(-1,0){2}}
\put(24,32){\line(-1,0){2}}
\put(48,32){\line(-1,0){2}}
\put(72,32){\line(-1,0){2}}

\put(0,12){\line(-1,0){2}}
\put(24,12){\line(-1,0){2}}
\put(48,12){\line(-1,0){2}}
\put(72,12){\line(-1,0){2}}

\put(14,4){\line(0,1){2}}
\put(38,4){\line(0,1){2}}
\put(62,4){\line(0,1){2}} 


 \linethickness{0.66mm}

\put(14,44){\line(0,1){2}}
\put(38,44){\line(0,1){2}}
\put(62,44){\line(0,1){2}}

 \put(14,24){\line(0,1){2}}
\put(38,24){\line(0,1){2}}
\put(62,24){\line(0,1){2}}

 \put(14,64){\line(0,1){2}}
\put(38,64){\line(0,1){2}}
\put(62,64){\line(0,1){2}}

 
\put(4,62){\line(1,0){8}}        
\put(16,62){\line(1,0){8}}
\put(28,62){\line(1,0){8}}       
\put(40,62){\line(1,0){8}}
\put(52,62){\line(1,0){8}}       
\put(64,62){\line(1,0){8}}
\put(4,42){\line(1,0){8}}
\put(16,42){\line(1,0){8}}
\put(28,42){\line(1,0){8}}
\put(40,42){\line(1,0){8}}
\put(52,42){\line(1,0){8}}
\put(64,42){\line(1,0){8}}
\put(4,22){\line(1,0){8}}
\put(16,22){\line(1,0){8}}
\put(28,22){\line(1,0){8}}
\put(40,22){\line(1,0){8}}
\put(52,22){\line(1,0){8}}
\put(64,22){\line(1,0){8}}
%
%
\end{picture}
\vspace{3.0ex}
\caption{\label{fig:2DGridDPart3}%
A partitioning of variables in the dual Forney factor graph of the 2D Ising model. The thick edges represent $\tilde \X_A$ and the remaining thin
edges represent $\tilde \X_B$. 
}
\end{figure}
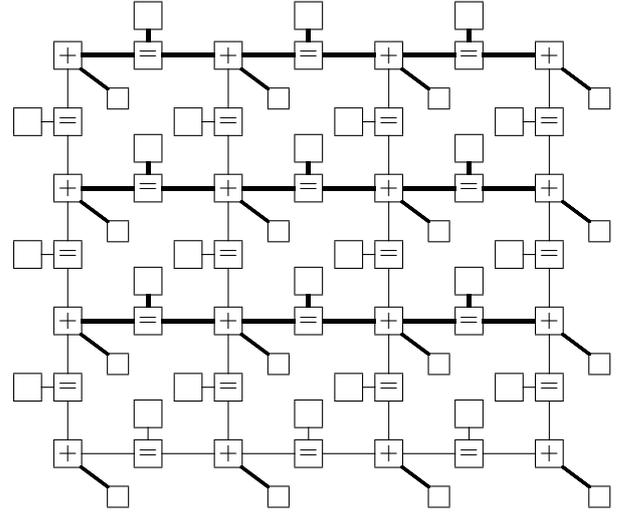

An example of such a partitioning is shown in Fig.~\ref{fig:2DGridDPart3}, where 
$\tilde \X_A$ is the set of all the variables associated with the thick 
edges and $\tilde \X_B$ the
set of all the variables associated with the remaining thin edges.
Accordingly, let $\B_A \subset \B$ contain the indices of the bonds 
marked by thick
edges and $\B_B = \B - \B_A$. 

For a valid configuration $\tilde \x = (\tilde \x_A, \tilde \x_B)$, 
let $\tilde \x_A = (\tilde \y, \tilde \z)$, where
$\tilde \y$ contains all the thick edges attached to 
the small unlabeled boxes (involved in~(\ref{eqn:IsingKernelDual2})) 
and $\tilde \z$ contains
all the variables associated with the thick bonds 
(involved in~(\ref{eqn:IsingKernelDual})). 

We prove that $\text{w}_{\text{H}} (\tilde\y)$, the Hamming weight of $\tilde\y$, is
always even, where the Hamming weight of a vector is the number
of non-zero components of that vector~\cite{RJM:77}.

%
\begin{trivlist}
\item{\bf Lemma 1. } 
If $\tilde \x$ is a valid configuration in
the dual Forney factor graph, 
then $\text{w}_{\text{H}} (\tilde \y)$ is even.
\end{trivlist}

\begin{trivlist}
\item \emph{Proof. } 
We consider $c = \bigoplus_{t=1}^N \tilde y_{t}$ the component-wise 
XOR of $\tilde \y$.
Each XOR factor 
imposes the
constraint that all its incident variables sum to $0$ in GF($2$).
Each $\tilde y_t$ in $c$ can thus be expanded as the XOR of the corresponding 
variables associated with the bonds, furthermore, the variables on the bonds each appear twice in this 
expansion. Hence $c = 0$, i.e., $\text{w}_{\text{H}} (\tilde\y)$ is even.
\hfill$\blacksquare$
\end{trivlist}

Lemma $1$ implies that
$Z_\mathrm{d}$, and thus $Z$ itself, are invariant under the change of sign 
of $H_m$.
Indeed, regardless of the sign 
of $H_m$ (i.e., assigned to all positive or to all negative values) 
$\prod_{m = 1}^{N} \lambda _{m}(\tilde x_m)$ takes on the same positive 
value, cf.~(\ref{eqn:IsingKernelDual2}).

The importance sampling algorithm works as follows. To draw
$\tilde \x^{(\ell)}$ at each iteration $\ell$, we first 
draw $\tilde \x^{(\ell)}_A$ 
according to 
a suitably defined auxiliary probability mass function on the bonds (see~(\ref{eqn:AuxDist})). 
We then update $\tilde \x^{(\ell)}_B$ to create a valid configuration 
$\tilde \x^{(\ell)} = (\tilde \x^{(\ell)}_A, \tilde \x^{(\ell)}_B)$. Updating $\tilde \x^{(\ell)}_B$ at each iteration is 
easy as $\tilde \x_B$ is a 
linear combination of $\tilde \x_A$. 


Let us define
\begin{IEEEeqnarray}{r;C;l}
\Lambda(\tilde \x_B) & \eqdef & 
\prod_{k \in \B_B} \!\!\gamma _{k}(\tilde x_k) \label{eqn:PartL1} \\
\Psi(\tilde \x_A) & \eqdef & 
\prod_{k \in \B_A} \!\!\gamma _{k}(\tilde x_k)\prod_{m = 1}^{N} \lambda _{m}(\tilde x_m)  \label{eqn:PartG1}
\end{IEEEeqnarray}
and
\begin{IEEEeqnarray}{r;C;l}
q(\tilde \x_A) & \eqdef & \frac{\Psi(\tilde \x_A)}{Z_{q}}, \qquad \forall \, \tilde \x_A \in \calX^{|\B_A|} \label{eqn:AuxDist}
\end{IEEEeqnarray}
where $Z_q$ in (\ref{eqn:AuxDist}) is available in closed form as
\begin{IEEEeqnarray}{r;C;l}
Z_{q}  & = & \sum_{\tilde \x_A} \Psi(\tilde \x_A) \\
        & = & 2^{|\B_A|}\, \text{exp}\big(\sum_{k \in \B_A} J_k - \sum_{m = 1}^{N} H_m\big) \label{eqn:Zq} 
\end{IEEEeqnarray}

Here $|\B_A|$ denotes the cardinality of $\B_A$. Note that in our set-up $H_m < 0$.
\vspace{0.05mm}



The product form of 
(\ref{eqn:PartG1}) suggests that 
to draw a
sample $\tilde \x_A^{(\ell)} = (\tilde \y^{(\ell)}, \tilde \z^{(\ell)})$ 
according to $q(\tilde \x_A)$, 
two separate 
subroutines are required, one subroutines for
the $\tilde \y^{(\ell)}$-part, 
and another one for the $\tilde \z^{(\ell)}$-part.

To draw the $\tilde \y^{(\ell)}$-part, we apply the following.
\vspace{0.1ex}

\begin{itemize}
\itemsep1.8pt
\item[] {\bf repeat}
\item[]  \hspace{4.5mm} \emph{draw} $u_1^{(\ell)}, u_2^{(\ell)}, \ldots, u_{N}^{(\ell)}\overset{\text{i.i.d.}}{\sim} \, \mathcal{U}[0,1]$
\item[]  \hspace{4.5mm} {\bf for} $m = 1$ {\bf to} $N$
\item [] \hspace{10mm} {\bf if} $u_m^{(\ell)} < \frac{1}{2}(1+e^{2H_m})$
\item [] \hspace{15.5mm}  $\tilde y_{m}^{(\ell)} = 0$
\item [] \hspace{10mm} {\bf else}
\item [] \hspace{15.5mm}  $\tilde y_{m}^{(\ell)} = 1$
\item [] \hspace{10mm} {\bf end if}
\item[]  \hspace{4.5mm} {\bf end for}
\item[] {\bf until} $\text{w}_{\text{H}}(\tilde\y^{(\ell)})$ \emph{is even}
\end{itemize}

The criteria to accept $\tilde\y^{(\ell)}$ is based on Lemma 1.
The quantity $\frac{1}{2}(1+ \,e^{2H_m})$ is equal to
$\lambda_m(0)/\big(\lambda_m(0) \,+\, \lambda_m(1)\big)$.

To draw the $\tilde \z^{(\ell)}$-part, the following subroutine is applied. 
\vspace{0.3ex}

\begin{itemize}
\itemsep1.7pt
\item[] \emph{draw} $u_1^{(\ell)}, u_2^{(\ell)}, \ldots, u_{|\B_A|}^{(\ell)}\overset{\text{i.i.d.}}{\sim} \,\mathcal{U}[0,1]$
\item[] {\bf for} $k = 1$ {\bf to} $|\B_A|$
\item [] \hspace{4.5mm} {\bf if} $u_k^{(\ell)} < \frac{1}{2}(1+e^{-2J_k})$
\item [] \hspace{10mm}  $\tilde z_{k}^{(\ell)} = 0$
\item [] \hspace{4.5mm} {\bf else}
\item [] \hspace{10mm}  $\tilde z_{k}^{(\ell)} = 1$ 
\item [] \hspace{4.5mm} {\bf end if}
\item[] {\bf end for}
\end{itemize}
\vspace{0.3ex}

Here, 
$\frac{1}{2}(1+ \,e^{-2J_k})$ is equal to
$\gamma_k(0)/\big(\gamma_k(0) \,+\, \gamma_k(1)\big)$.
We can then create $\tilde \x_A^{(\ell)}$ as a concatenation of 
$\tilde\y^{(\ell)}$ and $\tilde\z^{(\ell)}$. 

It is possible to compute the probability of rejection in  the algorithm. E.g., if the model is 
in a constant external field $H$
\begin{IEEEeqnarray}{r,C,l}
\text{P}\big(\text{w}_{\text{H}} (\tilde\y)\, \text{is odd}\big) &=&  \sinh(N|H|)e^{-N|H|} \\ 
											& \le & 0.5
\end{IEEEeqnarray}


The two previous subroutines will provide 
\emph{i.i.d.} samples $\tilde \x_A^{(1)}, \tilde \x_A^{(2)}, \ldots, \tilde \x_A^{(\ell)}, \ldots$ according 
to~(\ref{eqn:AuxDist}). 
Updating $\tilde \x_B^{(\ell)}$ is easy after generating $\tilde \x_A^{(\ell)}$. 
The created samples are then used in the following importance sampling algorithm in order to estimate $Z_{\mathrm{d}}$.
\vspace{0.3ex}

\begin{itemize}
\itemsep1.8pt
\item[] {\bf for} $\ell = 1$ {\bf to} $L$
\item[] \hspace{4.5mm} \emph{draw} $\x_A^{(\ell)}$ \emph{according to} $q(\tilde \x_A)$
\item[] \hspace{4.5mm} \emph{update} $\tilde \x_B^{(\ell)}$
\item[] {\bf end for}
\item[] \emph{compute}
\begin{IEEEeqnarray}{r,C,l}
\label{eqn:EstR}
\hat Z_{\text{IS}} & = & \frac{Z_q}{L} \sum_{\ell = 1}^L \Lambda(\tilde \x_B^{(\ell)}) 
\end{IEEEeqnarray}
\end{itemize}
\vspace{0.3ex}

\begin{trivlist}
\item{\bf Lemma 2. } 
$\hat Z_{\text{IS}}$ is an unbiased estimator 
of $Z_\mathrm{d}$. 
\end{trivlist}

\begin{trivlist}
\item \emph{Proof.} 
\vspace{-4mm}
\begin{IEEEeqnarray}{r;C;l}
\E_q[\, \hat Z_{\text{IS}}\,] & = & \E_q\Big[\, \frac{Z_q}{L} \sum_{\ell = 1}^L \Lambda(\tilde \X_B^{(\ell)}) \,\Big] \notag \\ 
						   & = & Z_q\cdot\E_q\big[\,\Lambda(\tilde \X_B)\,] \label{eqn:ZqExp} \notag \\
						   & = & \sum_{\tilde \x_A} \Psi(\tilde \x_A)\cdot\Lambda(\tilde \x_B) \notag \\
						   & = & Z_\mathrm{d} \notag
\end{IEEEeqnarray} 
\hfill$\blacksquare$
\end{trivlist}

The estimate of $Z_\mathrm{d}$ is then used to compute a 
Monte Carlo estimate of
$Z$, as in~(\ref{eqn:PartFunction}), via the normal factor 
graph duality theorem (cf.\ Section~\ref{sec:NFGD}).


The accuracy of~(\ref{eqn:EstR}) depends on the
fluctuations of $\Lambda(\tilde \x_B)$. If $\Lambda(\tilde \x_B)$ varies 
smoothly, $\hat Z_{\text{IS}}$ will
have a small variance. 
From (\ref{eqn:IsingKernelDual}) and~(\ref{eqn:PartL1}), 
we expect to observe a small variance if $J_k$ is large 
for $k \in \B_B$ -- as for large values of $J_k$, 
each factor~(\ref{eqn:IsingKernelDual}) tends to a constant factor. 
For more details, see~\cite{MeMo:2014b}.

\begin{figure}[t]
\setlength{\unitlength}{0.88mm}
\centering
\begin{picture}(77,71.6)(0,0)
\small
\put(0,60){\framebox(4,4){$+$}}
 \put(8,54){\framebox(3,3){}}
\put(12,60){\framebox(4,4){$=$}}
\put(24,60){\framebox(4,4){$+$}}
 \put(32,54){\framebox(3,3){}}
\put(36,60){\framebox(4,4){$=$}}
\put(48,60){\framebox(4,4){$+$}}
 \put(56,54){\framebox(3,3){}}
\put(60,60){\framebox(4,4){$=$}}
\put(72,60){\framebox(4,4){$+$}}
 \put(80,54){\framebox(3,3){}}
 \put(0,50){\framebox(4,4){$=$}}
\put(24,50){\framebox(4,4){$=$}}
\put(48,50){\framebox(4,4){$=$}}
\put(72,50){\framebox(4,4){$=$}}
\put(0,40){\framebox(4,4){$+$}}
 \put(8,34){\framebox(3,3){}}
\put(12,40){\framebox(4,4){$=$}}
\put(24,40){\framebox(4,4){$+$}}
 \put(32,34){\framebox(3,3){}}
\put(36,40){\framebox(4,4){$=$}}
\put(48,40){\framebox(4,4){$+$}}
 \put(56,34){\framebox(3,3){}}
\put(60,40){\framebox(4,4){$=$}}
\put(72,40){\framebox(4,4){$+$}}
 \put(80,34){\framebox(3,3){}}
\put(0,30){\framebox(4,4){$=$}}
\put(24,30){\framebox(4,4){$=$}}
\put(48,30){\framebox(4,4){$=$}}
\put(72,30){\framebox(4,4){$=$}}
\put(0,20){\framebox(4,4){$+$}}
 \put(8,14){\framebox(3,3){}}
\put(12,20){\framebox(4,4){$=$}}
\put(24,20){\framebox(4,4){$+$}}
 \put(32,14){\framebox(3,3){}}
\put(36,20){\framebox(4,4){$=$}}
\put(48,20){\framebox(4,4){$+$}}
 \put(56,14){\framebox(3,3){}}
\put(60,20){\framebox(4,4){$=$}}
\put(72,20){\framebox(4,4){$+$}}
 \put(80,14){\framebox(3,3){}}
\put(0,10){\framebox(4,4){$=$}}
\put(24,10){\framebox(4,4){$=$}}
\put(48,10){\framebox(4,4){$=$}}
\put(72,10){\framebox(4,4){$=$}}
\put(0,0){\framebox(4,4){$+$}}
 \put(8,-6){\framebox(3,3){}}
\put(12,0){\framebox(4,4){$=$}}
\put(24,0){\framebox(4,4){$+$}}
\put(32,-6){\framebox(3,3){}}
\put(36,0){\framebox(4,4){$=$}}
\put(48,0){\framebox(4,4){$+$}}
 \put(56,-6){\framebox(3,3){}}
\put(60,0){\framebox(4,4){$=$}}
\put(72,0){\framebox(4,4){$+$}}
 \put(80,-6){\framebox(3,3){}}

\put(12,66){\framebox(4,4){$$}}
\put(36,66){\framebox(4,4){$$}}
\put(60,66){\framebox(4,4){$$}}
\put(12,46){\framebox(4,4){$$}}
\put(36,46){\framebox(4,4){$$}}
\put(60,46){\framebox(4,4){$$}}
\put(12,26){\framebox(4,4){$$}}
\put(36,26){\framebox(4,4){$$}}
\put(60,26){\framebox(4,4){$$}}
\put(12,6){\framebox(4,4){$$}}
\put(36,6){\framebox(4,4){$$}}
\put(60,6){\framebox(4,4){$$}}
\put(-6,50){\framebox(4,4){$$}}
\put(18,50){\framebox(4,4){$$}}
\put(42,50){\framebox(4,4){$$}}
\put(66,50){\framebox(4,4){$$}}
\put(-6,30){\framebox(4,4){$$}}
\put(18,30){\framebox(4,4){$$}}
\put(42,30){\framebox(4,4){$$}}
\put(66,30){\framebox(4,4){$$}}
\put(-6,10){\framebox(4,4){$$}}
\put(18,10){\framebox(4,4){$$}}
\put(42,10){\framebox(4,4){$$}}
\put(66,10){\framebox(4,4){$$}}

\put(4,60){\line(4,-3){4}}
\put(4,60){\line(4,-3){4}}
\put(4.1,59.9){\line(4,-3){4}}
\put(4.1,60.1){\line(4,-3){4}}
\put(3.9,59.9){\line(4,-3){4}}
\put(28,60){\line(4,-3){4}}
\put(28,60){\line(4,-3){4}}
\put(28.1,59.9){\line(4,-3){4}}
\put(28.1,60.1){\line(4,-3){4}}
\put(27.9,59.9){\line(4,-3){4}}
 \put(52,60){\line(4,-3){4}}
 \put(52,60){\line(4,-3){4}}
 \put(52.1,59.9){\line(4,-3){4}}
 \put(52.1,60.1){\line(4,-3){4}}
 \put(51.9,59.9){\line(4,-3){4}}
 \put(76,60){\line(4,-3){4}}
 \put(76,60){\line(4,-3){4}}
 \put(76.1,59.9){\line(4,-3){4}}
 \put(76.1,60.1){\line(4,-3){4}}
 \put(75.9,59.9){\line(4,-3){4}}
 \put(4,40){\line(4,-3){4}}
 \put(4,40){\line(4,-3){4}}
 \put(4.1,39.9){\line(4,-3){4}}
 \put(4.1,40.1){\line(4,-3){4}}
 \put(3.9,39.9){\line(4,-3){4}}
 \put(28,40){\line(4,-3){4}}
 \put(28,40){\line(4,-3){4}}
 \put(28.1,39.9){\line(4,-3){4}}
 \put(28.1,40.1){\line(4,-3){4}}
 \put(27.9,39.9){\line(4,-3){4}}
 \put(52,40){\line(4,-3){4}}
 \put(52,40){\line(4,-3){4}}
 \put(52.1,39.9){\line(4,-3){4}}
 \put(52.1,40.1){\line(4,-3){4}}
 \put(51.9,39.9){\line(4,-3){4}}
 \put(76,40){\line(4,-3){4}}
 \put(76,40){\line(4,-3){4}}
 \put(76.1,39.9){\line(4,-3){4}}
 \put(76.1,40.1){\line(4,-3){4}}
 \put(75.9,39.9){\line(4,-3){4}}
 \put(4,20){\line(4,-3){4}}
 \put(4,20){\line(4,-3){4}}
 \put(4.1,19.9){\line(4,-3){4}}
 \put(4.1,20.1){\line(4,-3){4}}
 \put(3.9,19.9){\line(4,-3){4}}
 \put(28,20){\line(4,-3){4}}
 \put(28,20){\line(4,-3){4}}
 \put(28.1,19.9){\line(4,-3){4}}
 \put(28.1,20.1){\line(4,-3){4}}
 \put(27.9,19.9){\line(4,-3){4}}
 \put(52,20){\line(4,-3){4}}
 \put(52,20){\line(4,-3){4}}
 \put(52.1,19.9){\line(4,-3){4}}
 \put(52.1,20.1){\line(4,-3){4}}
 \put(51.9,19.9){\line(4,-3){4}}
 \put(76,20){\line(4,-3){4}}
 \put(76,20){\line(4,-3){4}}
 \put(76.1,19.9){\line(4,-3){4}}
 \put(76.1,20.1){\line(4,-3){4}}
 \put(75.9,19.9){\line(4,-3){4}}
 \put(4,0){\line(4,-3){4}}
 \put(4,0){\line(4,-3){4}}
 \put(4.1,-0.1){\line(4,-3){4}}
 \put(4.1,0.1){\line(4,-3){4}}
 \put(3.9,-0.1){\line(4,-3){4}}
 \put(28,0){\line(4,-3){4}}
 \put(28,0){\line(4,-3){4}}
 \put(28.1,-0.1){\line(4,-3){4}}
 \put(28.1,0.1){\line(4,-3){4}}
 \put(27.9,-0.1){\line(4,-3){4}}
 \put(52,0){\line(4,-3){4}}
 \put(52,0){\line(4,-3){4}}
 \put(52.1,-0.1){\line(4,-3){4}}
 \put(52.1,0.1){\line(4,-3){4}}
 \put(51.9,-0.1){\line(4,-3){4}}
 \put(76,0){\line(4,-3){4}}
 \put(76,0){\line(4,-3){4}}
 \put(76.1,-0.1){\line(4,-3){4}}
 \put(76.1,0.1){\line(4,-3){4}}
 \put(75.9,-0.1){\line(4,-3){4}}
\put(4,60){\line(4,-3){4}}
\put(4,60){\line(4,-3){4}}
\put(4.1,59.9){\line(4,-3){4}}
\put(4.1,60.1){\line(4,-3){4}}
\put(3.9,59.9){\line(4,-3){4}}
\put(28,60){\line(4,-3){4}}
\put(28,60){\line(4,-3){4}}
\put(28.1,59.9){\line(4,-3){4}}
\put(28.1,60.1){\line(4,-3){4}}
\put(27.9,59.9){\line(4,-3){4}}
 \put(52,60){\line(4,-3){4}}
 \put(52,60){\line(4,-3){4}}
 \put(52.1,59.9){\line(4,-3){4}}
 \put(52.1,60.1){\line(4,-3){4}}
 \put(51.9,59.9){\line(4,-3){4}}
 \put(76,60){\line(4,-3){4}}
 \put(76,60){\line(4,-3){4}}
 \put(76.1,59.9){\line(4,-3){4}}
 \put(76.1,60.1){\line(4,-3){4}}
 \put(75.9,59.9){\line(4,-3){4}}
 \put(4,40){\line(4,-3){4}}
 \put(4,40){\line(4,-3){4}}
 \put(4.1,39.9){\line(4,-3){4}}
 \put(4.1,40.1){\line(4,-3){4}}
 \put(3.9,39.9){\line(4,-3){4}}
 \put(28,40){\line(4,-3){4}}
 \put(28,40){\line(4,-3){4}}
 \put(28.1,39.9){\line(4,-3){4}}
 \put(28.1,40.1){\line(4,-3){4}}
 \put(27.9,39.9){\line(4,-3){4}}
 \put(52,40){\line(4,-3){4}}
 \put(52,40){\line(4,-3){4}}
 \put(52.1,39.9){\line(4,-3){4}}
 \put(52.1,40.1){\line(4,-3){4}}
 \put(51.9,39.9){\line(4,-3){4}}
 \put(76,40){\line(4,-3){4}}
 \put(76,40){\line(4,-3){4}}
 \put(76.1,39.9){\line(4,-3){4}}
 \put(76.1,40.1){\line(4,-3){4}}
 \put(75.9,39.9){\line(4,-3){4}}
 \put(4,20){\line(4,-3){4}}
 \put(4,20){\line(4,-3){4}}
 \put(4.1,19.9){\line(4,-3){4}}
 \put(4.1,20.1){\line(4,-3){4}}
 \put(3.9,19.9){\line(4,-3){4}}
 \put(28,20){\line(4,-3){4}}
 \put(28,20){\line(4,-3){4}}
 \put(28.1,19.9){\line(4,-3){4}}
 \put(28.1,20.1){\line(4,-3){4}}
 \put(27.9,19.9){\line(4,-3){4}}
 \put(52,20){\line(4,-3){4}}
 \put(52,20){\line(4,-3){4}}
 \put(52.1,19.9){\line(4,-3){4}}
 \put(52.1,20.1){\line(4,-3){4}}
 \put(51.9,19.9){\line(4,-3){4}}
 \put(76,20){\line(4,-3){4}}
 \put(76,20){\line(4,-3){4}}
 \put(76.1,19.9){\line(4,-3){4}}
 \put(76.1,20.1){\line(4,-3){4}}
 \put(75.9,19.9){\line(4,-3){4}}
 \put(4,0){\line(4,-3){4}}
 \put(4,0){\line(4,-3){4}}
 \put(4.1,-0.1){\line(4,-3){4}}
 \put(4.1,0.1){\line(4,-3){4}}
 \put(3.9,-0.1){\line(4,-3){4}}
 \put(28,0){\line(4,-3){4}}
 \put(28,0){\line(4,-3){4}}
 \put(28.1,-0.1){\line(4,-3){4}}
 \put(28.1,0.1){\line(4,-3){4}}
 \put(27.9,-0.1){\line(4,-3){4}}
 \put(52,0){\line(4,-3){4}}
 \put(52,0){\line(4,-3){4}}
 \put(52.1,-0.1){\line(4,-3){4}}
 \put(52.1,0.1){\line(4,-3){4}}
 \put(51.9,-0.1){\line(4,-3){4}}
 \put(76,0){\line(4,-3){4}}
 \put(76,0){\line(4,-3){4}}
 \put(76.1,-0.1){\line(4,-3){4}}
 \put(76.1,0.1){\line(4,-3){4}}
 \put(75.9,-0.1){\line(4,-3){4}}
%
\put(2,54){\line(0,1){6}}
\put(2,50){\line(0,-1){6}}
\put(74,54){\line(0,1){6}}
\put(74,50){\line(0,-1){6}}
\put(2,34){\line(0,1){6}}
\put(2,30){\line(0,-1){6}}
\put(26,34){\line(0,1){6}}
\put(26,30){\line(0,-1){6}}
\put(50,34){\line(0,1){6}}
\put(50,30){\line(0,-1){6}}
\put(74,34){\line(0,1){6}}
\put(74,30){\line(0,-1){6}}
\put(2,14){\line(0,1){6}}
\put(2,10){\line(0,-1){6}}
\put(74,14){\line(0,1){6}}
\put(74,10){\line(0,-1){6}}
\put(4,2){\line(1,0){8}}
\put(16,2){\line(1,0){8}}
\put(52,2){\line(1,0){8}}
\put(64,2){\line(1,0){8}}
\put(0,52){\line(-1,0){2}}
\put(72,52){\line(-1,0){2}}

\put(0,32){\line(-1,0){2}}
\put(24,32){\line(-1,0){2}}
\put(48,32){\line(-1,0){2}}
\put(72,32){\line(-1,0){2}}

\put(0,12){\line(-1,0){2}}
\put(72,12){\line(-1,0){2}}

\put(14,4){\line(0,1){2}}
\put(62,4){\line(0,1){2}} 

\put(14,64){\line(0,1){2}}
\put(62,64){\line(0,1){2}}

\put(4,62){\line(1,0){8}}        
\put(16,62){\line(1,0){8}}
\put(52,62){\line(1,0){8}}       
\put(64,62){\line(1,0){8}}

\put(28,42){\line(1,0){8}}
\put(40,42){\line(1,0){8}}

\put(38,44){\line(0,1){2}}

\put(4,22){\line(1,0){8}}
\put(16,22){\line(1,0){8}}

\put(52,22){\line(1,0){8}}
\put(64,22){\line(1,0){8}}

\put(14,24){\line(0,1){2}}
\put(62,24){\line(0,1){2}}

 \linethickness{0.65mm}
 
 \put(26,14){\line(0,1){6}}
\put(26,10){\line(0,-1){6}}
\put(50,14){\line(0,1){6}}
\put(50,10){\line(0,-1){6}}
\put(24,12){\line(-1,0){2}}
\put(48,12){\line(-1,0){2}}

\put(28,2){\line(1,0){8}}
\put(40,2){\line(1,0){8}}

\put(14,44){\line(0,1){2}}
\put(62,44){\line(0,1){2}}

\put(38,24){\line(0,1){2}}

\put(38,64){\line(0,1){2}}

\put(38,4){\line(0,1){2}}

\put(26,54){\line(0,1){6}}
\put(26,50){\line(0,-1){6}}

\put(50,54){\line(0,1){6}}
\put(50,50){\line(0,-1){6}}

\put(24,52){\line(-1,0){2}}
\put(48,52){\line(-1,0){2}}

 
\put(28,62){\line(1,0){8}}       
\put(40,62){\line(1,0){8}}
\put(4,42){\line(1,0){8}}
\put(16,42){\line(1,0){8}}
\put(52,42){\line(1,0){8}}
\put(64,42){\line(1,0){8}}
\put(28,22){\line(1,0){8}}
\put(40,22){\line(1,0){8}}
%
%
\end{picture}
\vspace{3.5ex}
\caption{\label{fig:2DGridDAnother}%
Another example of a partitioning of variables in the dual Forney factor graph of the 2D Ising model. 
}
\end{figure}
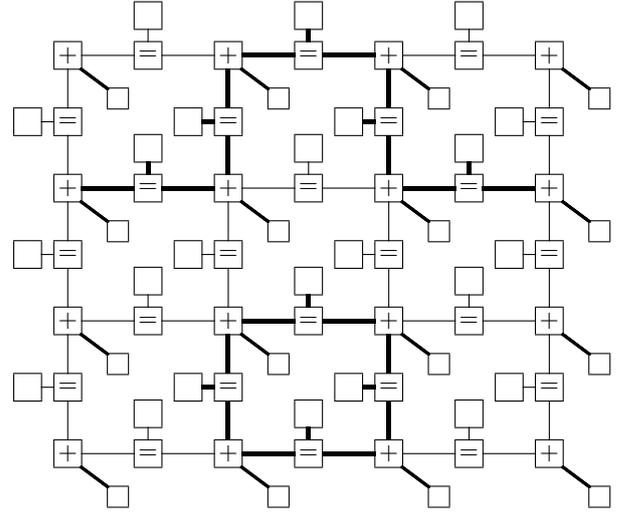

We emphasize that our choice of partitioning
in Fig.~\ref{fig:2DGridDPart3} 
is not unique. 
Fig.~\ref{fig:2DGridDAnother} shows another example of a partitioning
in the dual Forney factor graph whose corresponding partitioning in
the primal factor graph is not cycle-free. 
A partitioning which gives rise to a slightly different importance sampling algorithm (with no 
rejections) is discussed in~\cite{MeMo:2014b}.

The proposed algorithm is applicable to the Ising model in the absence of
an external field as well. 
Indeed, partitionings in Figs.~\ref{fig:2DGridDPart3} and~\ref{fig:2DGridDAnother} 
are valid even when the external field is not present.
We will consider Ising models without an external field
in our numerical experiments in Section~\ref{sec:NumIsing2}.

That being the case, to observe fast convergence in the dual domain,
not all the coupling parameters need to be strong, but a restricted 
subset of them. The method of this paper can thus 
be regarded as supplementary to 
the ones presented in~\cite{MoLo:ISIT2013} and~\cite{MeMo:2014a}, where the focus
is on models at low temperature (corresponding to models in which all the
coupling parameters are strong) and on models in a strong external field. 

\section{Numerical Experiments}
\label{sec:Num}


We apply the importance sampling algorithm 
to estimate the log partition function 
per site, i.e., $\frac{1}{N}\ln Z$, of 
2D Ising models.
All simulation results show $\frac{1}{N}\ln Z$ vs.\ the number
of samples for \emph{one instance}\footnote{In statistical physics, estimating quantities for a fixed set of
couplings (generated according to some distribution) is called the ``quenched
average".} of the model with periodic
boundaries.

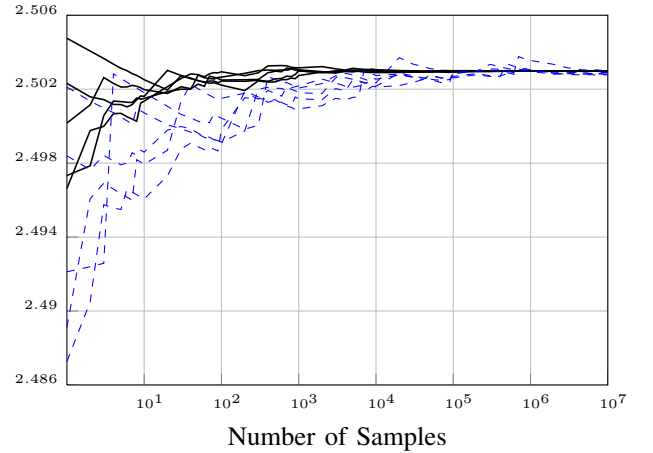
\begin{figure}[h!!!!]
\centering
\begin{tikzpicture}
\begin{axis}[
			height = 41.0ex,
			width = 55.0ex,
			grid = major,
			tick pos=left, 
			xmode=log,
			xminorticks = false,	
		    yminorticks = false,	
		    y tick label style={
        /pgf/number format/.cd,
            fixed,
            precision=3,
        /tikz/.cd
    		}, 				
		xtick={1e1, 1e2, 1e3, 1e4, 1e5, 1e6, 1e7},
	    ytick={2.486, 2.49, 2.494, 2.498, 2.502, 2.506},
		xlabel= Number of Samples, style={font=\normalsize},
			xmin = 1,
			xmax = 1e7,
			ymax = 2.506,
			ymin = 2.486,
			yticklabel style = {font=\tiny,yshift=0.5ex},
            xticklabel style = {font=\tiny,xshift=0.5ex}			
			]
\pgfplotstableread{./ZS1.txt}\mydataone
\pgfplotstableread{./ZS2.txt}\mydatatwo
\pgfplotstableread{./ZS3.txt}\mydatathree
\pgfplotstableread{./ZS4.txt}\mydatafour
\pgfplotstableread{./ZS5.txt}\mydatafive
\pgfplotstableread{./ZU1.txt}\mydatasix
\pgfplotstableread{./ZU2.txt}\mydataseven
\pgfplotstableread{./ZU3.txt}\mydataeight
\pgfplotstableread{./ZU4.txt}\mydatanine
\pgfplotstableread{./ZU5.txt}\mydataten
		\addplot [
		 color = black,
		 semithick
		]		
		 table[y = Z] from \mydataone;
		 
		 \addplot [
 		 color = blue,
 		 dashed,
 		 thin
		 ]
 		  table[y = Z] from \mydatasix;
 		 
 		 \addplot [
 		 color = black,
 		 semithick
 		 ]
 		  table[y = Z] from \mydatatwo;

		 \addplot [
 		 color = black,
		 semithick
 		 ]
 		  table[y = Z] from \mydatathree;
 		  
 		 \addplot [
 		 color = black,
 		 semithick
		 ]
 		  table[y = Z] from \mydatafour;
 		  
 		 \addplot [
 		 color = black,
 		 		 semithick
		 ]
 		  table[y = Z] from \mydatafive;

   		 \addplot [
 		 color = blue,
 		  		 dashed
		 ]
 		  table[y = Z] from \mydataseven;
 		  
   		 \addplot [
 		 color = blue,
  		 dashed
		 ]
 		  table[y = Z] from \mydataeight;
 		  
   		 \addplot [
 		 color = blue,
  		 dashed
		 ]
 		  table[y = Z] from \mydatanine;
 		  
   		 \addplot [
 		 color = blue,
  		 dashed
		 ]
 		  table[y = Z] from \mydataten;

\end{axis}
\end{tikzpicture}
\caption{\label{fig:Fer1}%
Estimated log partition function per site vs.\ the number of samples
for a $30\times 30$ Ising model, with $J_k\sim\calU[1.0, 1.25]$ for $k \in \B_A$ 
and $J_k\sim\calU[1.25, 1.5]$ for $k \in \B_B$. 
The plot shows five different sample paths obtained from importance 
sampling (solid black lines) and five different sample paths obtained from uniform 
sampling (dashed blue lines) on the dual factor graph.}
  \end{figure}



We consider 2D ferromagnetic Ising models with spatially 
varying (edge-dependent) coupling 
parameters without an external field in 
Section~\ref{sec:NumIsing2}
We will also compare the efficiency of  
the importance sampling algorithm with
uniform sampling.  
Comparisons with Gibbs sampling 
and the Swendsen-Wang 
algorithm~\cite{SW:87} 
are discussed in~\cite{MeMo:2014b}.

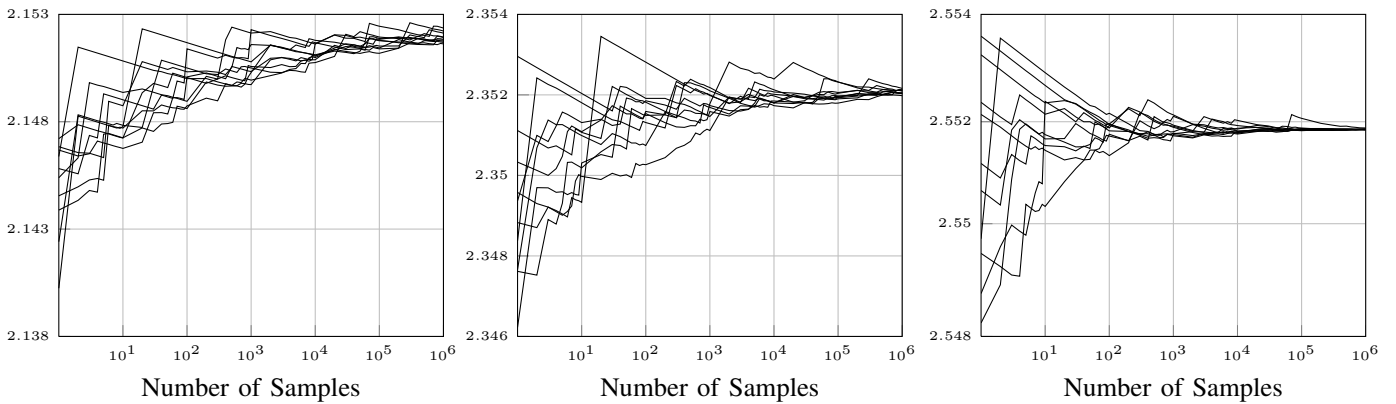
\begin{figure*}[t!!!]
\centering
\begin{subfigure}[b]{0.28\textwidth}
\centering
\hspace*{-8.5ex}
\begin{tikzpicture}
\begin{axis}[
			height = 37.0ex,
      	    width = 42.0ex,
			grid = major,
			tick pos=left, 
			xmode=log,
			xminorticks = false,	
		    yminorticks = false,	
		    y tick label style={
        /pgf/number format/.cd,
            fixed,
            precision=3,
        /tikz/.cd
    		}, 				
		xtick={1e1, 1e2, 1e3, 1e4, 1e5, 1e6},
			ytick={2.138, 2.143, 2.148, 2.153},
		xlabel= Number of Samples = {font=\normalsize},
			xmin = 1,
			xmax = 1e6,
			ymin = 2.138,
			ymax = 2.153,
			yticklabel style = {font=\tiny,yshift=0.05ex},
            xticklabel style = {font=\tiny,xshift=0.0ex}			
			]
\pgfplotstableread{./Z41.txt}\mydataone
\pgfplotstableread{./Z42.txt}\mydatatwo
\pgfplotstableread{./Z44.txt}\mydatafour
\pgfplotstableread{./Z45.txt}\mydatafive
\pgfplotstableread{./Z46.txt}\mydatasix
\pgfplotstableread{./Z47.txt}\mydataseven
\pgfplotstableread{./Z48.txt}\mydataeight
\pgfplotstableread{./Z49.txt}\mydatanine
\pgfplotstableread{./Z410.txt}\mydataten
\pgfplotstableread{./Z411.txt}\mydataeleven

		\addplot [
		color = black,
		]		
    		table[y = Z] from \mydataone;
		 
 		 \addplot [
 		 color = black
 		 ]
 		  table[y = Z] from \mydatatwo;	 
 
 		  
 		 \addplot [
 		 color = black
 		 ]
 		  table[y = Z] from \mydatafour;	
 		  
 		 \addplot [
 		 color = black
 		 ]
 		  table[y = Z] from \mydatafive;	

 		 \addplot [
 		 color = black
 		 ]
 		  table[y = Z] from \mydatasix;	
 		  
 		   		 \addplot [
 		 color = black
 		 ]
 		  table[y = Z] from \mydataseven;	
 		  
 		   		 \addplot [
 		 color = black
 		 ]
 		  table[y = Z] from \mydataeight;	
		   		 
		   		 \addplot [
 		 color = black
 		 ]
 		  table[y = Z] from \mydatanine;	
 		  
 		   		 \addplot [
 		 color = black
 		 ]
 		  table[y = Z] from \mydataten;	

		   		 \addplot [
 		 color = black
 		 ]
 		  table[y = Z] from \mydataeleven;

\end{axis}
\end{tikzpicture}
\end{subfigure}
\hspace{-2.8ex}
\begin{subfigure}[b]{0.28\textwidth}
\centering
\begin{tikzpicture}
\begin{axis}[
			height = 37.0ex,
			width = 42.0ex,
			grid = major,
			tick pos=left, 
			xmode=log,
			xminorticks = false,	
		    yminorticks = false,	
		    y tick label style={
        /pgf/number format/.cd,
            fixed,
            precision=3,
        /tikz/.cd
    		}, 				
		xtick={1e1, 1e2, 1e3, 1e4, 1e5, 1e6},
			ytick={2.346, 2.348, 2.350, 2.352, 2.354},
		xlabel= Number of Samples = {font=\normalsize},
			xmin = 1,
			xmax = 1e6,
			ymin = 2.346,
			ymax = 2.354,
			yticklabel style = {font=\tiny,yshift=0.05ex},
            xticklabel style = {font=\tiny,xshift=0.0ex}			
			]
\pgfplotstableread{./Z51.txt}\mydataone
\pgfplotstableread{./Z52.txt}\mydatatwo
\pgfplotstableread{./Z53.txt}\mydatathree
\pgfplotstableread{./Z54.txt}\mydatafour
\pgfplotstableread{./Z55.txt}\mydatafive
\pgfplotstableread{./Z56.txt}\mydatasix
\pgfplotstableread{./Z57.txt}\mydataseven
\pgfplotstableread{./Z58.txt}\mydataeight
\pgfplotstableread{./Z59.txt}\mydatanine
\pgfplotstableread{./Z510.txt}\mydataten

		\addplot [
				 color = black
		 ]		
		 table[y = Z] from \mydataone;
		 
 		 \addplot [
 		 color = black
 		 ]
 		  table[y = Z] from \mydatatwo;	 
 
  		 \addplot [
 		 color = black
 		 ]
 		  table[y = Z] from \mydatathree;	
 		  
 		 \addplot [
 		 color = black
 		 ]
 		  table[y = Z] from \mydatafour;	
 		  
 		 \addplot [
 		 color = black
 		 ]
 		  table[y = Z] from \mydatafive;	

 		 \addplot [
 		 color = black
 		 ]
 		  table[y = Z] from \mydatasix;	
 		  
 		   		 \addplot [
 		 color = black
 		 ]
 		  table[y = Z] from \mydataseven;	
 		  
 		   		 \addplot [
 		 color = black
 		 ]
 		  table[y = Z] from \mydataeight;	

  		 \addplot [
 		 color = black
 		 ]
 		  table[y = Z] from \mydatanine;	

  		 \addplot [
 		 color = black
 		 ]
 		  table[y = Z] from \mydataten;	 		  
 
\end{axis}
\end{tikzpicture}
  \end{subfigure}
\hspace{4.9ex}
\begin{subfigure}[b]{0.28\textwidth}
\centering
\begin{tikzpicture}
\begin{axis}[
			height = 37.0ex,
			width = 42.0ex,
			grid = major,
			tick pos=left, 
			xmode=log,
			xminorticks = false,	
		    yminorticks = false,	
		    y tick label style={
        /pgf/number format/.cd,
            fixed,
            precision=3,
        /tikz/.cd
    		}, 				
		xtick={1e1, 1e2, 1e3, 1e4, 1e5, 1e6},
			ytick={2.548, 2.5501, 2.552, 2.554},
		xlabel= Number of Samples = {font=\small},
			xmin = 1,
			xmax = 1e6,
			ymin = 2.548,
			ymax = 2.554,
			yticklabel style = {font=\tiny,yshift=0.05ex},
            xticklabel style = {font=\tiny,xshift=0.0ex}		
			]
\pgfplotstableread{./Z61.txt}\mydataone
\pgfplotstableread{./Z62.txt}\mydatatwo
\pgfplotstableread{./Z63.txt}\mydatathree
\pgfplotstableread{./Z64.txt}\mydatafour
\pgfplotstableread{./Z65.txt}\mydatafive
\pgfplotstableread{./Z66.txt}\mydatasix
\pgfplotstableread{./Z67.txt}\mydataseven
\pgfplotstableread{./Z68.txt}\mydataeight
\pgfplotstableread{./Z69.txt}\mydatanine
\pgfplotstableread{./Z610.txt}\mydataten

		\addplot [
		 color = black
		]		
		 table[y = Z] from \mydataone;
		 
 		 \addplot [
 		 color = black
 		 ]
 		  table[y = Z] from \mydatatwo;	 
 
  		 \addplot [
 		 color = black
 		 ]
 		  table[y = Z] from \mydatathree;	
 		  
 		 \addplot [
 		 color = black
 		 ]
 		  table[y = Z] from \mydatafour;	
 		  
 		 \addplot [
 		 color = black
 		 ]
 		  table[y = Z] from \mydatafive;	
 		  
  		 \addplot [
 		 color = black
 		 ]
 		  table[y = Z] from \mydatasix;	
 		  
 		 \addplot [
 		 color = black
 		 ]
 		  table[y = Z] from \mydataseven;	
 		  
		 \addplot [
 		 color = black
 		 ]
 		  table[y = Z] from \mydataeight;	

		 \addplot [
 		 color = black
 		 ]
 		  table[y = Z] from \mydatanine;	
 		  
 		 \addplot [
 		 color = black
 		 ]
 		  table[y = Z] from \mydataten;	

\end{axis}
\end{tikzpicture}
 \end{subfigure}
\caption{Estimated log partition function per site vs.\ the number of samples
for a $50\times 50$ Ising model, with
$J_k\sim\calU[0.1, 1.0]$ for $k \in \B_A$ and $H_m \sim\calU[-0.8, -0.2]$ for $1 \le m \le N$; 
for $k \in \B_B$
(left) $J_k\sim\calU[1.0, 1.2]$, (middle) $J_k\sim\calU[1.2, 1.4]$, and (right) $J_k\sim\calU[1.4, 1.6]$. Each plot shows ten different sample paths obtained from importance sampling on the dual factor graph.}
\label{fig:FerExtern1}
\end{figure*}

The 2D ferromagnetic Ising models in an external field 
with spatially varying model parameters are considered in Section~\ref{sec:NumIsing1}.

\subsection{2D Ising models without an external field} 
\label{sec:NumIsing2}

We consider a 2D Ising model of size $N = 30\times 30$ without 
an external magnetic field. For $k \in \B_A$, we 
set $J_k\overset{\text{i.i.d.}}{\sim}\calU[1.0, 1.25]$ and
for $k \in \B_B$, 
set $J_k\overset{\text{i.i.d.}}{\sim}\calU[1.25, 1.5]$.

Fig.~\ref{fig:Fer1} shows simulation
results obtained from importance sampling (solid lines) and from uniform 
sampling (dashed lines) in the dual Forney factor graph. 
From Fig.~\ref{fig:Fer1}, the estimated log partition function per site 
is about $2.503$.

We observe that importance sampling outperforms 
uniform sampling (with virtually the same amount of 
computation time). For more details, see also~\cite{MeMo:2014a, MeMo:2014b}.


\subsection{2D Ising models in an external field} 
\label{sec:NumIsing1}

We set $N = 50\times 50$, $J_{k} \overset{\text{i.i.d.}}{\sim} \calU[0.1, 1.0]$ for $k \in \B_A$, and 
$H_m\overset{\text{i.i.d.}}\sim\calU[-0.8, -0.2]$ for $1 \le m \le N$ in all the experiments.

In the first experiment, $J_{k} \overset{\text{i.i.d.}}{\sim} \calU[1.0, 1.2]$
for $k \in \B_B$.
Simulation results obtained from importance sampling in the dual 
factor graph are 
shown in Fig.~\ref{fig:FerExtern1} (left). 
In the second experiment, $J_{k} \overset{\text{i.i.d.}}{\sim} \calU[1.4, 1.5]$
for $k \in \B_B$. 
Fig.~\ref{fig:FerExtern1} (middle) shows simulation results.
We set $J_{k} \overset{\text{i.i.d.}}{\sim} \calU[1.4, 1.6]$
for $k \in \B_B$ in the third experiment. 
Simulation results
are shown in Fig.~\ref{fig:FerExtern1} (right), where
the estimated $\frac{1}{N}\ln Z$ is about $2.5518$. 
Notice that in Fig.~\ref{fig:FerExtern1} from left to right, the range of the $y$-axis is 0.015, 0.008, and 0.006, respectively.

In agreement with our analysis in Section~\ref{sec:IS}, 
we observe that convergence improves as $J_k$ becomes larger for $k \in \B_B$.

\section{Conclusion}

An importance sampling algorithm was presented for estimating the partition function 
of the 2D ferromagnetic Ising model in a consistent external 
magnetic field. The algorithm is described in 
the dual Forney factor graph representing the model. After introducing 
a partitioning and an auxiliary importance sampling 
distribution, the method operates by first simulating a subset
of the variables, followed by doing computations 
over the remaining ones.
The algorithm can efficiently estimate the partition function 
when the model is at low temperature or when the model 
contains a 
mixture of strong and weak coupling parameters. 
The proposed algorithm is applicable to the 3D Ising model and the
$q$-state Potts model in an external field
as well.
For duality results in the context of statistical 
physics, see, e.g.,~\cite{KW:41, Savit:80},~\cite[Chapter 10]{NO:11}.

\section*{Acknowledgements}

The author would like to thank Hans-Andrea Loeliger, David 
Forney, and Justin Dauwels for 
their helpful comments. 
The author would also like to thank Pascal Vontobel 
for proofreading an earlier 
version of this paper and for pointing out to him~\cite{Savit:80}.

\newcommand{\IT}{IEEE Trans.\ Inf.\ Theory}
\newcommand{\CASI}{IEEE Trans.\ Circuits \& Systems~I}
\newcommand{\COM}{IEEE Trans.\ Comm.}
\newcommand{\COMLet}{IEEE Commun.\ Lett.}
\newcommand{\COMMag}{IEEE Communications Mag.}
\newcommand{\ETT}{Europ.\ Trans.\ Telecomm.}
\newcommand{\SPMag}{IEEE Signal Proc.\ Mag.}
\newcommand{\ProcIEEE}{Proceedings of the IEEE}

\end{document}